\documentclass[sigplan,nonacm,10pt]{acmart}

\renewcommand\footnotetextcopyrightpermission[1]{}
\AtBeginDocument{%
  }

\usepackage{microtype}
\microtypesetup{protrusion=false}
\usepackage{graphicx}
\usepackage{placeins}
\usepackage{subcaption}
\usepackage{booktabs}
\usepackage{amsmath}
\usepackage{mathtools}
\usepackage{enumitem}
\usepackage{tabularx}
\usepackage[ruled,vlined]{algorithm2e}
\usepackage[capitalize,noabbrev]{cleveref}
\crefname{algocf}{Algorithm}{Algorithms}
\Crefname{algocf}{Algorithm}{Algorithms}
\usepackage{tikz}
\usepackage{pgfplots}
\usetikzlibrary{arrows.meta,backgrounds,calc,fit,patterns,positioning,shapes.geometric,shapes.misc}
\pgfplotsset{compat=1.18}
\graphicspath{{figures/}}
\setcopyright{acmlicensed}
\copyrightyear{2026}
\acmYear{2026}
\acmDOI{XXXXXXX.XXXXXXX}
\acmConference[ATC '26]{The 2026 ACM SIGOPS Annual Technical Conference}{November 16--18, 2026}{Hong Kong, China}
\acmISBN{978-1-4503-XXXX-X/2026/04}

\definecolor{TEEColor}{RGB}{52, 140, 102}
\definecolor{FHEColor}{RGB}{106, 76, 147}
\definecolor{PDColor}{RGB}{34, 110, 178}
\definecolor{UnifiedColor}{RGB}{211, 110, 59}
\definecolor{EncodeColor}{RGB}{191, 134, 33}
\definecolor{BoundaryGray}{RGB}{85, 85, 85}

\newcommand{\ct}{\mathsf{ct}}

\newcommand{\pk}{\mathsf{pk}}
\newcommand{\sk}{\mathsf{sk}}
\newcommand{\Enc}{\mathsf{Enc}}
\newcommand{\Dec}{\mathsf{Dec}}

\newcommand{\policycpu}{CPU TEE}

\newcommand{\policypd}{\texorpdfstring{Bifrost\textsuperscript{+}}{Bifrost+}}
\newcommand{\policyunified}{Bifrost}
\newcommand{\policyfhe}{Pure FHE}
\newcommand{\runhead}[1]{\par\smallskip\noindent\textbf{#1.}\ }

\setlist[itemize]{leftmargin=1.5em}
\SetAlFnt{\footnotesize}
\SetAlCapFnt{\small}
\SetAlCapNameFnt{\small}

\begin{document}

\title{Bifrost: Hybrid TEE--FHE Inference for Privacy-Preserving Transformer and LLM Serving}

\author{Chenghao Chen}
\email{ch.chen@sjtu.edu.cn}
\affiliation{%
  \institution{Shanghai Jiao Tong University}
  \city{Shanghai}
  \country{China}}

\author{Kailun Qin}
\email{kailun.qin@sjtu.edu.cn}
\affiliation{%
  \institution{Shanghai Jiao Tong University}
  \city{Shanghai}
  \country{China}}

\author{Xiaolin Zhang}
\email{xiaolinzhang@sjtu.edu.cn}
\affiliation{%
  \institution{Shanghai Jiao Tong University}
  \city{Shanghai}
  \country{China}}

\author{Chi Zhang}
\email{zcsjtu@sjtu.edu.cn}
\affiliation{%
  \institution{Shanghai Jiao Tong University}
  \city{Shanghai}
  \country{China}}

\author{Dawu Gu}
\email{dwgu@sjtu.edu.cn}
\affiliation{%
  \institution{Shanghai Jiao Tong University}
  \city{Shanghai}
  \country{China}}

\renewcommand{\shortauthors}{Chen et al.}

\begin{abstract}
Cloud-hosted transformer and large language model (LLM) inference creates an immediate confidentiality problem: user prompts may contain source code, business plans, personal data, or regulated documents, yet remote serving exposes intermediate state to the cloud software stack and accelerator runtime. Fully homomorphic encryption (FHE) keeps accelerator-side execution ciphertext-only, but end-to-end LLM inference remains expensive because linear layers are interleaved with non-linear, reduction-heavy, cache-state, and refresh-sensitive operators. CPU trusted execution environments (TEEs) can execute those operators natively, but trusting a CPU TEE does not by itself explain how an untrusted accelerator should participate.

We present \policyunified{}, a hybrid TEE--FHE serving architecture for this trust-policy point: secrets are provisioned only to an attested CPU TEE, while the accelerator, device memory, driver/runtime stack, and host software remain outside the trusted computing base. FHE is therefore the secure delegation mechanism, not the sole trust anchor. \policyunified{} keeps regular projection and feed-forward linear layers on accelerator-backed CKKS, while non-linear operators, attention-side control logic, KV-state transitions, and decrypt-then-encrypt (DtE) refresh execute inside the CPU TEE.

\policypd{} further applies a prefill/decode (PD) split to the same boundary. Instead of sending every prompt token through the encrypted accelerator loop, \policypd{} builds prompt-side KV state inside the CPU TEE and hands only decode-side state to the hybrid ciphertext path. This does not make FHE kernels faster; it avoids prompt-side encrypted work when the CPU TEE can execute prefill directly.

In an estimator-style comparison matching Euston's methodology, \policyunified{} reduces projected latency by 9.25$\times$ on GPT-2 (1.5B) and 9.91$\times$ on LLaMA~3 (8B). In direct CKKS/FHE deployments, \policypd{} reduces TTFT by 14.6--45.8$\times$ on GPT-2 (124M) and 15.3--53.4$\times$ on Qwen3 (0.6B) for prompt lengths 16 and 64. Cost breakdowns show the remaining boundary: fully cached GPT-2 decode is GEMM-dominated, while Qwen3 exposes a 34.4\% fallback weight-encoding penalty when encoded weights no longer fully fit on the accelerator. The resulting lesson is selective encrypted execution: use FHE only where ciphertext-only accelerator delegation is required, and keep non-linear, refresh, and prompt-side work inside the CPU TEE.
\end{abstract}

\ccsdesc[500]{Security and privacy~Systems security}
\ccsdesc[300]{Computer systems organization~Architectures}

\keywords{fully homomorphic encryption, trusted execution environment, large language model, privacy-preserving inference, heterogeneous secure execution}

\maketitle
\pagestyle{plain}

\section{Introduction}

Large language models (LLMs) are increasingly deployed through Model-as-a-Service (MaaS): users send prompts to a cloud provider, the provider hosts the model, and generated outputs are returned remotely. This deployment model is convenient and cost-effective, but it creates a direct privacy problem. Prompts may contain source code, business plans, personal information, legal text, or regulated documents, and remote inference exposes intermediate state to the cloud software stack, accelerator runtime, and service operator.

Privacy-enhancing technologies offer two natural starting points. Fully Homomorphic Encryption (FHE) protects data cryptographically by enabling computation on ciphertexts \cite{gentryFullyHomomorphicEncryption2009a, cheonHomomorphicEncryptionArithmetic2017}. Trusted Execution Environments (TEEs), such as Intel TDX and AMD SEV, protect data architecturally by confining plaintext execution to hardware-isolated domains \cite{IntelTrustDomaina, AMDSecureEncrypted}. For LLM serving, however, neither mechanism is sufficient by itself. FHE provides a narrow accelerator-side privacy boundary, but non-linear operators, depth growth, ciphertext refresh, and serving state make end-to-end inference expensive. CPU TEEs execute those operators natively, but CPU TEE-only execution becomes the trusted-CPU latency floor rather than a way to use untrusted accelerators.

This paper targets the remaining deployment point. A tenant provisions secrets to an attested CPU TEE, but does not place accelerator memory, firmware, drivers, or runtime in the trusted computing base. FHE is therefore a secure accelerator-delegation mechanism rather than the trust root. The question is not whether FHE beats plaintext CPU TEE execution; it does not in our prototype. The question is which operators and serving phases should still pay FHE cost when accelerator execution must remain ciphertext-only.

Two properties of modern LLM serving make this question more than encrypted GEMM scheduling. First, model architectures evolve: decoder blocks combine normalization, RoPE, gating, grouped-query attention, and cache logic whose semantics vary across model families and are awkward to re-engineer as fixed homomorphic circuits. Second, serving is phase-structured: prefill determines time-to-first-token (TTFT), while decode repeatedly touches persistent KV state and small per-token control paths. A confidential serving system therefore needs both an operator-level placement policy and a phase-level serving policy.

Recent CKKS systems, compilers, and accelerators have improved encrypted linear algebra \cite{gaoEustonEfficientUserFriendlyCorrected, zhangSecureTransformerInference2025a, pmlr-v267-de-castro25a, fanWarpDriveGPUBasedFully2025a, jiaoNeoEfficientFully2025, cryptoeprint:2024/1543}; hybrid systems such as BLB, TEEFHE, and HT2ML show that selected computation can leave a pure-cryptographic path when the protection model allows it \cite{blb25usenixsecurity, wangScalableFullyHomomorphic2019, wangHT2MLEfficientHybrid2023a}. What remains missing, to our knowledge, is a TEE+FHE Transformer/LLM serving system that couples a trusted CPU to accelerator-backed FHE, supports autoregressive cache semantics, respects prefill/decode asymmetry, and evaluates end-to-end behavior.

The key observation behind \policyunified{} is that the trust boundary and the Transformer layer boundary should not coincide. Regular linear projections are stable, data-parallel, and FHE-friendly, so they are the right candidates for the ciphertext accelerator path. Non-linear operators, control logic, refresh, and serving-state transitions are precision-sensitive or control-heavy, so they should execute inside the CPU TEE when the trust policy allows it. This split adapts to changing model architectures while keeping the accelerator view ciphertext-only.

\begin{figure}[htbp]
\centering
\includegraphics[width=\columnwidth]{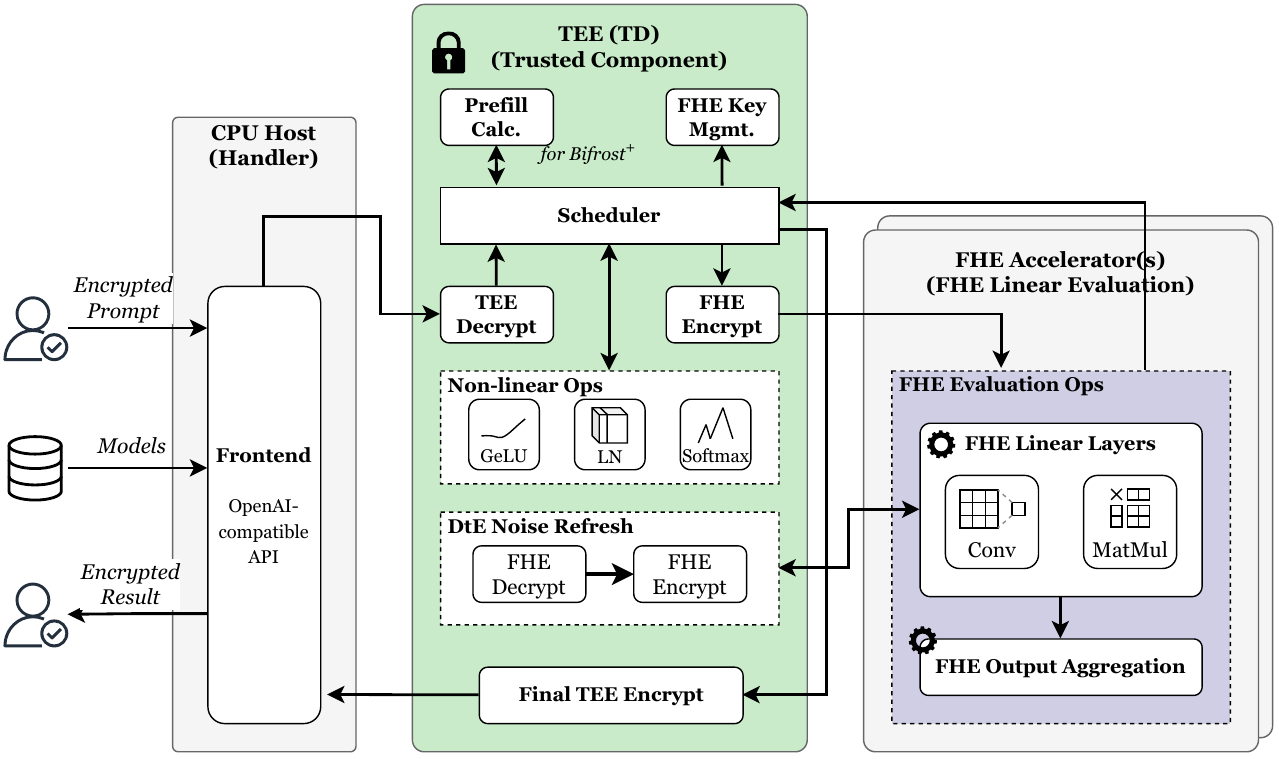}
\caption{End-to-end hybrid inference architecture. The CPU TEE is the trusted control point; the accelerator is untrusted and evaluates FHE linear layers over ciphertexts. Trusted GPU execution is not assumed.}
\Description{End-to-end system overview showing the client, CPU TEE, and an untrusted FHE accelerator connected through ciphertext-only transfers.}
\label{fig:system_overview}
\end{figure}

\policyunified{} instantiates this observation as a hybrid TEE--FHE architecture. Large linear layers run on an untrusted accelerator under CKKS, while normalization, RoPE, Softmax-style reductions, attention-side control, gating, and ciphertext refresh execute inside the CPU TEE through decrypt-compute-encrypt or decrypt-then-encrypt (DtE) windows. The prototype is built as an executable serving runtime, not a collection of isolated kernels: it includes request routing, policy control, encoded-weight caching, and KV-cache machinery for autoregressive inference. \Cref{fig:system_overview} summarizes the end-to-end path.

\policyunified{} alone still routes prompt prefill and steady-state decode through the same hybrid path. That is a poor serving policy because every prompt token pays the encrypted accelerator cost before the first token can be emitted. \policypd{} applies prefill/decode (PD) disaggregation \cite{zhongDistServeDisaggregatingPrefill2024, patelSplitwiseEfficientGenerative2024} to the TEE--FHE boundary. It completes prompt prefill and prompt-side KV construction inside the CPU TEE, then hands only decode-side state to the same hybrid ciphertext path. This is not a new cryptographic primitive and does not accelerate the FHE kernels; it removes prompt-side encrypted work.

At the implementation level, our prototype builds on nano-vLLM, a lightweight derivative of the vLLM serving stack \cite{kwonEfficientMemoryManagement2023, yuGeeeekExplorerNanovllm2026}. This lets the evaluation exercise an end-to-end serving substrate, including request routing, policy selection, and KV state, rather than relying only on operator microbenchmarks or replay scripts.

\runhead{Contributions}
We make four contributions.
\begin{itemize}
    \item \textbf{CPU-TEE-rooted ciphertext-only accelerator delegation.} We define and instantiate a deployment point where the CPU TEE is the only trusted execution boundary and the accelerator acts only as a ciphertext evaluator.
    \item \textbf{Operator-affinity TEE--FHE execution.} \policyunified{} keeps regular linear layers on accelerator-backed FHE and moves non-linear, reduction-heavy, control-heavy, and refresh-sensitive operators into the CPU TEE.
    \item \textbf{Phase-aware PD split over a secure serving boundary.} \policypd{} keeps prompt prefill and prompt-side KV construction inside the TEE, then hands only decode-side state to the hybrid ciphertext path.
    \item \textbf{Executable serving substrate and deployment-boundary evidence.} The nano-vLLM-derived prototype integrates policy routing, encoded-weight caching, and KV backends, exposing encoded-weight residency and accelerator memory pressure---not arithmetic alone---as the dominant scaling boundary for larger deployments.
\end{itemize}

\section{Background and Motivation}
\label{sec:background}

\subsection{Fully Homomorphic Encryption}
\label{sec:background-fhe}

Our prototype targets the CKKS approximate homomorphic encryption scheme \cite{cheonHomomorphicEncryptionArithmetic2017}. CKKS operates over the ring $R_Q = \mathbb{Z}_Q[X]/(X^N + 1)$ for power-of-two ring dimension $N$, and packs $N/2$ complex slots into each ciphertext. This packing makes CKKS well suited to large linear algebra because one ciphertext can hold many vector or matrix fragments at once. The price is that scale management, key switching, and rotations are explicit costs rather than transparent implementation details.

Three CKKS concepts are central for this paper. First, \textbf{ring dimension} controls both security and throughput: larger $N$ exposes more slots, but also increases ciphertext size and the cost of NTT-heavy kernels. Second, \textbf{levels} and \textbf{rescaling} govern multiplicative depth: after a ciphertext--ciphertext multiplication, CKKS rescales to keep the numerical scale bounded while consuming one level from the modulus chain. Third, \textbf{bootstrapping} refreshes a ciphertext once the remaining level budget is exhausted, but that refresh is expensive enough to be a first-order systems concern.

This is why operator choice matters. Let $D_{\mathrm{lin}}$ denote the depth contribution from linear layers and $D_{\mathrm{NL}}$ the contribution from non-linear approximations. In a pure-FHE transformer block, the effective encrypted depth is
\begin{equation}
D_{\mathrm{FHE}} \approx D_{\mathrm{lin}} + D_{\mathrm{NL}}.
\end{equation}
For modern LLMs, $D_{\mathrm{NL}}$ is often the more fragile term because normalization, Softmax-style reductions, and gating paths must be approximated or otherwise lowered into expensive arithmetic circuits. Once those operators move into a trusted plaintext domain, the encrypted depth becomes
\begin{equation}
D_{\mathrm{FHE}}^{\mathrm{hybrid}} \approx D_{\mathrm{lin}},
\end{equation}
which directly reduces refresh pressure and allows the FHE path to focus on accelerator-friendly linear algebra.

\subsection{Trusted Execution Environments}
\label{sec:background-tee}

TEEs protect computation architecturally rather than cryptographically. In Intel TDX, guest-private memory is encrypted, access control is enforced by the CPU memory controller, and attestation binds the running software stack to platform measurements before secrets are provisioned \cite{IntelTrustDomaina}. For our purposes, TDX offers three useful properties: plaintext execution of complex operators, an attestation root for key release, and a deployment model that does not require trusting the host kernel or hypervisor.

Those benefits come with systems constraints. TEE performance is still shaped by trusted-memory capacity, page-management overhead, and the limited availability of accelerator-class TEEs for general-purpose inference. AMD SEV offers a similar VM-level protection story with different isolation and attestation details \cite{AMDSecureEncrypted}. Recent work on confidential LLM inference across CPU and GPU TEEs and surveys of heterogeneous confidential computing quantify the performance, cost, and trust-model tradeoffs of bringing accelerators into the TCB \cite{chrapekConfidentialLLMInference2025,wangConfidentialComputingHeterogeneous2026}. In both cases, TEEs are much better at executing branch-heavy or reduction-heavy operators than FHE, but they do not remove the problem of serving large models inside a limited trusted footprint.

\subsection{Autoregressive Transformer Inference}
\label{sec:background-transformer}

This paper targets decoder-only autoregressive serving. Each block applies normalization, QKV projection, positional encoding, attention against the KV cache, output projection, another normalization, and feed-forward layers before producing logits for the next token. The same block structure is exercised during both prompt processing and generation, but the serving phases behave differently.

\textbf{Prefill} processes all prompt tokens, is matrix-heavy, and benefits from batch-style linear algebra. \textbf{Decode} emits one token at a time, touches the KV cache repeatedly, and performs smaller per-step computation. The KV cache is therefore not an optional side detail: it is central to latency and to the placement of persistent state across the TEE--accelerator boundary. Any secure serving system has to explain not only where arithmetic runs, but also how cache state is materialized, updated, and handed across phases.

\subsection{Design Space and Motivation}
\label{sec:background-motivation}

The relevant design space has four endpoints: CPU TEE-only, CPU+GPU TEE, pure FHE, and CPU TEE+FHE accelerator delegation. CPU+GPU TEE designs form a complementary trusted-accelerator endpoint \cite{chrapekConfidentialLLMInference2025,wangConfidentialComputingHeterogeneous2026}; Bifrost instead keeps the accelerator outside the TCB and uses FHE for ciphertext-only delegation. The systems opportunity exploited by this paper is the complementarity at the last point: large linear projections map naturally to accelerator-backed CKKS execution, while normalization, Softmax-style reductions, routing, and refresh are precisely the stages where FHE becomes depth-intensive or numerically fragile and where a TEE can execute the operator exactly.

Autoregressive serving introduces a second asymmetry on top of the operator split. Prompt prefill is matrix-heavy and batched, while decode performs smaller per-token work with frequent KV accesses. In plaintext serving, PD disaggregation separates these phases for throughput and latency reasons. In our hybrid setting, a unified path still forces every prompt token through the expensive encrypted loop, so TTFT grows with prompt length even if the decode path is already well designed. This is why the paper applies PD disaggregation to the TEE--FHE boundary as a \emph{PD split}: a second-layer trust-domain partitioning decision on top of the operator split, rather than a replacement for operator-affinity execution.

Finally, memory residency shapes how far the design scales. We use parameterized model names throughout: the common GPT-2 Small, Medium, and Large checkpoints correspond to GPT-2 (124M), GPT-2 (355M), and GPT-2 (774M) in our parameter-count convention \cite{gpt-2-radford2019language}. GPT-2 (124M) behaves like a clean fully cached case, GPT-2 (355M) remains close to that regime, and Qwen3 (0.6B), a dense model from the Qwen3 family \cite{yangQwen3TechnicalReport2025}, already pays a measurable fallback encoding penalty. Once encoded weights, persistent KV state, and temporary workspace compete for the same accelerator budget, memory becomes a first-order boundary even if the architectural split remains beneficial.

\section{System Design}
\label{sec:design}

Our goal is not to introduce a new primitive, but to present a systems-level execution strategy that makes privacy-preserving transformer and LLM serving more practical. This section focuses on operator placement, execution flow, scheduler policy, and threat model; Section~\ref{sec:impl} then maps those ideas onto the prototype runtime.

\subsection{Architecture Overview}
\label{sec:arch-overview}

\begin{figure*}[htbp]
\centering
\includegraphics[width=.9\textwidth,height=.72\textheight,keepaspectratio]{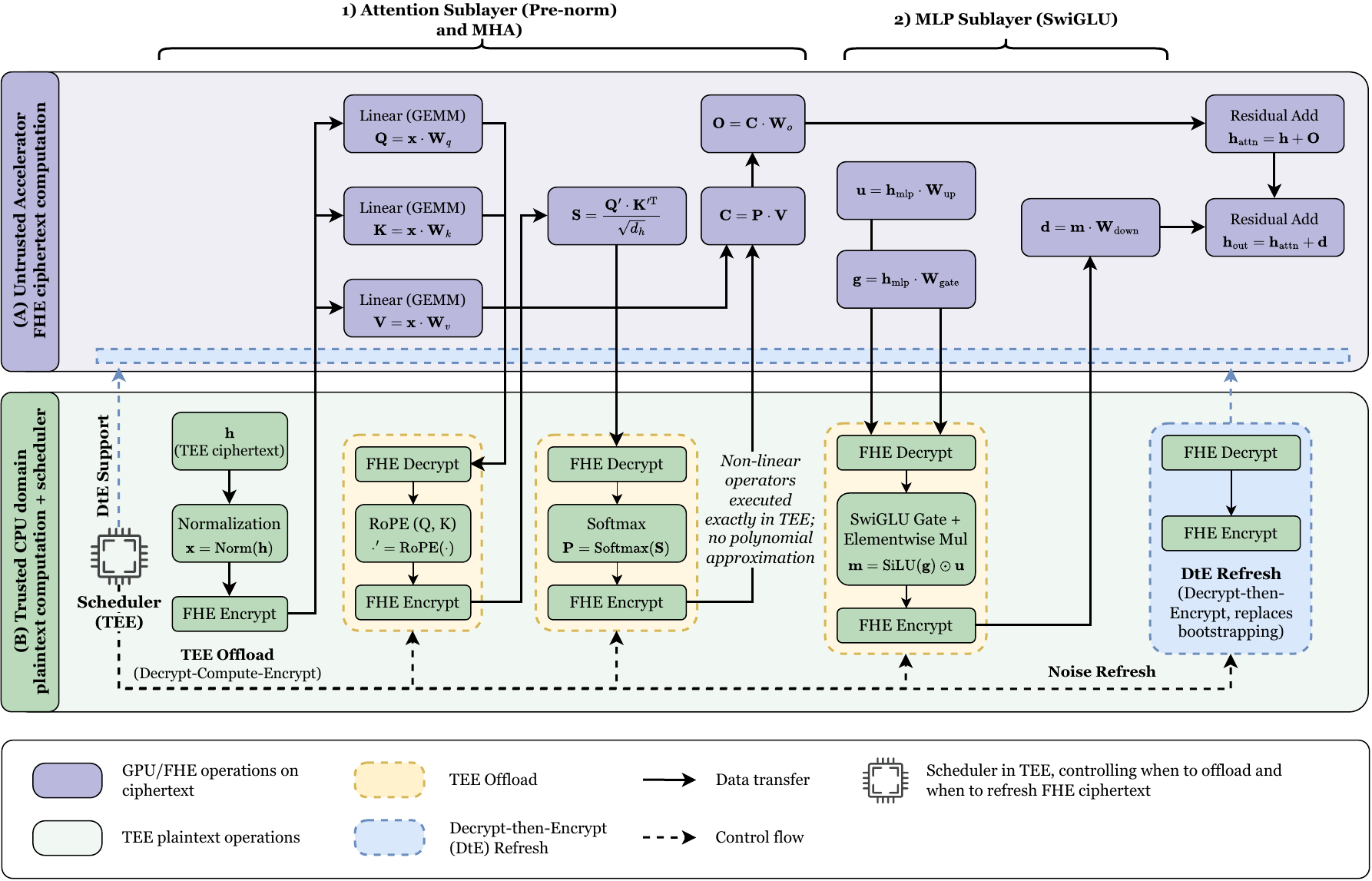}
\caption{\textbf{Operator-level hybrid dataflow of one representative modern LLM block.} Linear and accelerator-friendly segments run on an untrusted FHE accelerator ciphertext path, instantiated by a GPU in our prototype. Non-linear, control-heavy, and refresh-sensitive operators, such as normalization, RoPE, Softmax, gating, and DtE refresh, are handled inside the CPU TEE through decrypt-compute-encrypt offloads.}
\Description{Operator-level dataflow of a representative modern LLM block. The FHE accelerator evaluates linear ciphertext segments, while the CPU TEE handles representative non-linear operators and decrypt-then-encrypt refresh.}
\label{fig:dataflow}
\end{figure*}

\runhead{Operator-affinity split}
Our design partitions transformer inference by operator affinity, as illustrated in \cref{fig:dataflow}:
\begin{itemize}
    \item \textbf{FHE-friendly (accelerator).} Large, regular linear projections, including QKV/output projection GEMMs and FFN linear layers, are evaluated homomorphically on an accelerator-backed FHE path. These operators dominate FLOPs and remain structurally stable across model variants.
    \item \textbf{TEE-friendly (CPU).} Operators that are precision-fragile, depth-hungry, or control-heavy under FHE, including RoPE, Softmax, attention control, gating, normalization, and refresh, are executed inside the TEE in plaintext and their outputs are immediately re-encrypted before re-entering the FHE path.
\end{itemize}
This split keeps the encrypted backbone focused on throughput-critical linear layers while making the overall system more robust to evolving LLM motifs.

\runhead{End-to-end execution flow}
\Cref{fig:system_overview} shows the deployment pipeline. The client encrypts the prompt and sends ciphertexts to the server. A host-side handler forwards the request into the TEE, which manages secret keys, executes TEE-friendly operators, and orchestrates accelerator-backed execution for FHE-friendly linear layers. The same TEE-side runtime also maintains the instantiated KV cache and governs when ciphertext-backed state is handed to the accelerator path.

Execution alternates between TEE segments and accelerator-backed FHE segments. For an \textbf{accelerator segment}, the TEE prepares the required activations and dispatches ciphertexts to the accelerator, where homomorphic linear kernels run without accessing plaintext. For a \textbf{TEE segment}, the TEE decrypts only the minimal intermediate state needed for the next operator window, evaluates it in plaintext, and immediately re-encrypts the outputs before resuming accelerator execution. Throughout inference, the untrusted accelerator observes only ciphertexts and deterministic kernel launches.

\begin{table}[htbp]
\caption{Execution paths and measurement roles. CPU TEE-only is the trusted-CPU latency floor; Bifrost targets the stricter setting where delegated accelerator work remains ciphertext-only.}
\label{tab:paths}
\centering
\footnotesize
\begin{tabularx}{\columnwidth}{lcccX@{}}
\toprule
\textbf{Path} & \textbf{Prefill} & \textbf{Decode} & \textbf{Accel.} & \textbf{Role} \\
\midrule
CPU TEE-only & TEE & TEE & none & latency floor \\
\policyfhe{} & FHE & FHE & ct & projected baseline \\
\policyunified{} & hybrid & hybrid & ct & direct CKKS/FHE \\
\policypd{} & TEE & hybrid & ct & direct CKKS/FHE + PD \\
\bottomrule
\end{tabularx}
\end{table}

Within this architecture, \cref{tab:paths} separates the trusted-CPU endpoint from the three encrypted or delegated paths; ct denotes a ciphertext-only accelerator view. CPU TEE-only executes both phases inside the trusted CPU and serves as the latency floor under our CPU-TEE trust assumption. \policyfhe{} is the fully encrypted crypto baseline. \policyunified{} is the base hybrid architecture, which retains the operator split but still routes prompt prefill through the same hybrid loop used by decode. \policypd{} keeps the same hybrid decode path while moving prompt prefill into the CPU TEE. It is built directly on top of \policyunified{} rather than being a separate architecture.

\subsection{Operator-Affinity Execution}
\label{sec:offload}

Our offloading strategy follows a simple systems observation: in transformer inference, the arithmetic bulk lies in large linear algebra, whereas the fragile parts under FHE are non-linear and reduction-heavy operators together with ciphertext refresh. We therefore split execution by operator affinity: the accelerator-backed FHE path specializes in data-parallel linear layers, while the CPU TEE executes operators whose FHE realizations are depth-intensive, rotation-heavy, or accuracy-fragile.

\runhead{Offloading non-linear operators}
Since FHE natively supports additions and multiplications, non-linear operators in LLMs are typically implemented via polynomial or rational approximations, or via comparison-oriented arithmetic circuits. Even with sufficient CKKS precision, replacing $f(\cdot)$ by an approximation $\tilde{f}(\cdot)$ changes the model's semantics. Our design instead executes these operators inside the CPU TEE in plaintext: when reaching a non-linear operator, the encrypted activation is transferred to the TEE, decrypted with the secret key sealed in the TEE, evaluated exactly using the native operator implementation, and immediately re-encrypted before returning to the FHE pipeline. This has three practical effects:
\begin{enumerate}[leftmargin=1.5em]
    \item \textbf{No approximation gap.} The remaining numerical effects are those already inherent to CKKS encoding and quantization.
    \item \textbf{Depth stays linear-dominated.} The encrypted path spends its level budget primarily on the large linear kernels that benefit from accelerator execution.
    \item \textbf{Architecture adaptivity.} As LLM blocks evolve, new non-linear semantics can be absorbed by the TEE without redesigning approximation circuits or re-calibrating the homomorphic backbone.
\end{enumerate}

The key insight is that only the four linear projections per layer---QKV, output, gate-up, and down---are candidates for FHE offload, and these projections account for 97.5\% of per-token decode time in the fully cached regime (\cref{tab:operator-domain}). All remaining operators execute on the CPU TEE at negligible cost relative to the homomorphic GEMM.

\begin{table}[htbp]
\caption{Operator-to-domain mapping and measured decode-time fractions for GPT-2 (124M) in the fully cached regime.}
\label{tab:operator-domain}
\centering
\footnotesize
\begin{tabular}{llr}
\toprule
\textbf{Operation} & \textbf{Domain} & \textbf{\% decode} \\
\midrule
Linear (QKV, O, GateUp, Down) & GPU FHE & 97.5\% \\
DtE boundary crossing          & TEE $\leftrightarrow$ GPU & 2.3\% \\
Norm, RoPE, Softmax/control, activation & CPU TEE & $<$0.2\% \\
\bottomrule
\end{tabular}
\end{table}

\runhead{Decrypt-then-encrypt (DtE) refresh}
Bootstrapping is the standard mechanism to reset ciphertext noise in CKKS, but it is expensive due to large-scale transforms and modular arithmetic. Our system replaces homomorphic refresh with a TEE-assisted DtE primitive:
\begin{equation}
\ct' = \Enc_{\pk}\!\left(\Dec_{\sk}(\ct)\right),
\end{equation}
where decryption and re-encryption both occur inside the TEE. Conceptually, DtE provides the same logical effect as bootstrapping---resetting noise and restoring decryptability---but performs it using the TEE trust boundary rather than evaluating the refresh homomorphically.

DtE is beneficial when
\begin{equation}
T_{\mathrm{transfer}} + T_{\mathrm{Dec}} + T_{\mathrm{Enc}}
\;<\;
T_{\mathrm{refresh}}^{\mathrm{FHE}}.
\end{equation}
Offloading non-linear operators and DtE refresh reinforce each other: removing non-linear approximations slows depth growth, and when refresh is still required, DtE makes it cheaper.

\runhead{Managing domain-transition overhead}
CPU--accelerator transfers could in principle offset the gains from hybrid execution. Our design mitigates this in two ways. First, ciphertexts remain resident on the accelerator across consecutive linear layers whenever possible, so transitions occur only at architectural boundaries or refresh points. Second, the runtime batches adjacent TEE-resident operators into a single offload window when possible. As a result, the number of cross-domain transitions scales with operator groups rather than with every elementwise step.

\subsection{Boundary-Aware Scheduler and Tiling}
\label{sec:scheduler}

The scheduler is the control plane of the hybrid inference pipeline. It runs inside the TEE and coordinates operator routing, ciphertext-state bookkeeping, domain transitions, and noise management. Beyond routing, its key systems responsibility is to choose an execution granularity that balances accelerator throughput against TEE-boundary costs.

\begin{figure}[htbp]
\centering
\includegraphics[width=\linewidth]{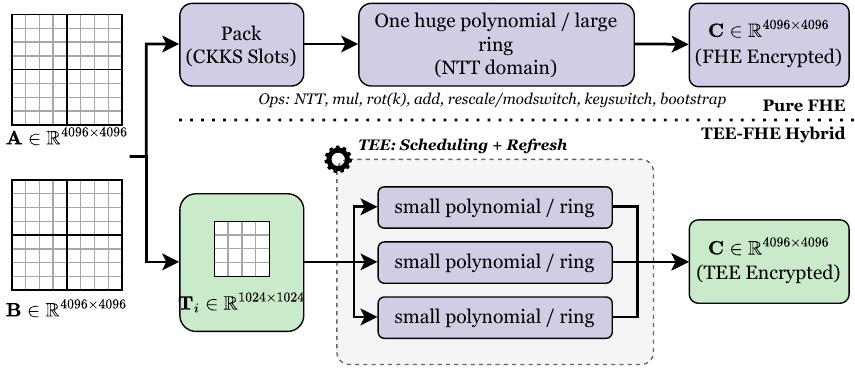}
\caption{\textbf{Boundary-aware scheduler for large homomorphic linear layers.} Instead of treating tiling as a GPU-only kernel decision, the scheduler jointly chooses tile size, persistent cache versus fallback layers, and refresh or repack insertion points. The resulting plan dispatches smaller CKKS tiles to the FHE accelerator while preserving the CPU TEE as the control point for cache and boundary management.}
\Description{Scheduler-controlled tiling diagram showing a large GEMM being decomposed into smaller tiles while the CPU TEE scheduler also reasons about cache fit, fallback encoding, and refresh insertion.}
\label{fig:scheduler}
\end{figure}

\begin{algorithm}[tbp]
\DontPrintSemicolon
\caption{Boundary-aware scheduler}
\label{alg:boundary-aware-scheduler}
\KwIn{operator graph $G$, tiling configurations $\mathcal{T}$, cache budget $B$}
\KwOut{execution plan $\Pi$}
$\Pi \leftarrow [\,]$\;
$b \leftarrow B$\;
\ForEach{operator group $g$ in topological order of $G$}{
  \uIf{$g$ is TEE-friendly}{
    append a TEE offload window for $g$ to $\Pi$\;
    update ciphertext level, scale, and boundary state\;
  }
  \Else{
    choose $\tau \in \mathcal{T}$ minimizing projected latency under budget $b$\;
    \uIf{encoded weights for $g$ fit in persistent accelerator cache}{
      pin $g$ in accelerator-resident cache\;
    }
    \Else{
      append fallback encode / repack action for $g$ to $\Pi$\;
    }
    append tiled FHE dispatch for $g$ with tile size $\tau$\;
  }
  \If{predicted level budget falls below threshold}{
    append DtE refresh to $\Pi$ and reset the noise state\;
  }
}
\Return{$\Pi$}\;
\end{algorithm}

\runhead{Tiling for large homomorphic GEMMs}
A monolithic CKKS packing strategy for large matrices often forces a large ring dimension to accommodate enough slots, which directly increases ciphertext size and the cost of NTT and key-switch-heavy kernels. Tiling replaces one large-ring evaluation with many small-ring evaluations: the scheduler cuts activations into tiles, packs them into smaller CKKS instances, and launches a sequence of tile-level kernels, as shown in \cref{fig:scheduler}. This improves accelerator efficiency and makes per-tile noise growth easier to track, at the cost of additional TEE-boundary crossings.

\runhead{Scheduling, assembly, and refresh}
Given a tiling plan, the scheduler dispatches tile-level GEMM tasks to the accelerator backend and receives encrypted assembled outputs by homomorphic aggregation. It maintains a lightweight ciphertext state, including remaining levels, scale, and noise margin, and inserts DtE refresh when the predicted budget is insufficient:
\[
\ct \;\rightarrow\; \Enc_{\pk}\!\bigl(\Dec_{\sk}(\ct)\bigr).
\]
Non-linear operators follow the same pattern: the scheduler transfers only the required ciphertext state into the TEE, executes the operator exactly in plaintext, and immediately re-encrypts the output before resuming FHE execution.

\runhead{Boundary-aware planning}
In practice, the scheduler optimizes more than raw operator order. It tries to maximize contiguous accelerator-resident linear segments, batch adjacent TEE-friendly operators into a single offload window, and avoid unnecessary re-encoding of weights or transient states. This is especially important for models whose encrypted decode path only partially fits in device memory. In those cases, the scheduler must jointly choose the tiling plan, which layers remain persistently cached on the accelerator, and whether the next boundary crossing should also be used to refresh or repack ciphertext state.

\subsection{From \texorpdfstring{\policyunified{}}{Bifrost} to \texorpdfstring{\policypd{}}{Bifrost+}}
\label{sec:pd-arch}

\begin{figure}[htbp]
\centering
\includegraphics[width=\linewidth]{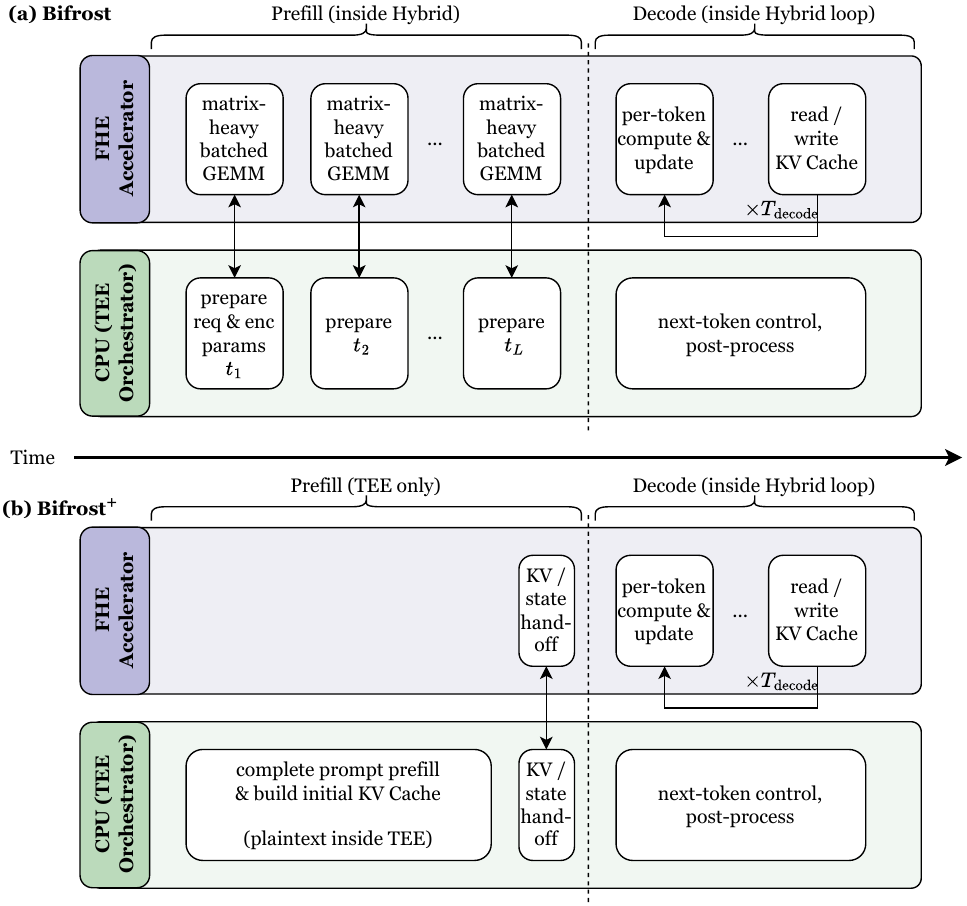}
\caption{\textbf{PD-oriented serving split over the hybrid backend.} \policyunified{} and \policypd{} share the same hybrid decode path. The difference is phase-specific: \policyunified{} routes every prompt token through the hybrid loop, whereas \policypd{} completes prompt prefill and prompt-side KV construction inside the CPU TEE before performing a single handoff into the same decode backend.}
\Description{A two-panel timeline comparing Bifrost and Bifrost plus. The left panel shows every prompt token traversing the hybrid loop, while the right panel shows TEE prefill, one KV handoff, and unchanged hybrid decode.}
\label{fig:pd-serving-split}
\end{figure}

Where the operator split (\cref{sec:offload}) partitions \emph{which operators} cross the trust boundary, the PD split partitions \emph{which serving phase} crosses it. Under \policyunified{}, prompt tokens and decode tokens both traverse the same hybrid loop, so the system materializes encrypted intermediate states and hybrid KV updates at every prompt step before the first token can be emitted. Under \policypd{}, prompt processing and prompt-side KV construction are completed inside the TEE, and only the decode-side state handoff enters the hybrid path. In other words, PD does not change the encrypted decode kernels; it changes which state reaches those kernels and when the handoff occurs. This adapts the PD disaggregation idea from plaintext serving infrastructure \cite{zhongDistServeDisaggregatingPrefill2024, patelSplitwiseEfficientGenerative2024} to the TEE--FHE trust boundary.

\par\smallskip
\noindent\textbf{Why not route prefill through FHE?}\par\nobreak
\noindent
The PD split is motivated by a simple cost asymmetry. Let $T$ be the prompt length, $c_{\mathrm{cpu}}$ the cost of one prompt-token step in plaintext inside the CPU TEE, and $c_{\mathrm{fhe}}$ the cost of one hybrid encrypted step. The three relevant policies have the following first-order TTFT behavior:
\begin{align}
\mathrm{TTFT}_{\mathrm{TEE}} &\approx (T+1)c_{\mathrm{cpu}}, \\
\mathrm{TTFT}_{\mathrm{Bifrost}} &\approx (T+1)c_{\mathrm{fhe}}, \\
\mathrm{TTFT}_{\mathrm{Bifrost}^{+}} &\approx T c_{\mathrm{cpu}} + c_{\mathrm{fhe}}.
\end{align}
In our prototype, $c_{\mathrm{cpu}} \ll c_{\mathrm{fhe}}$. CPU TEE-only is therefore expected to be faster than both \policyunified{} and \policypd{} under the CPU-TEE trust assumption. \policypd{} is not designed to beat that trusted-CPU endpoint. It is designed to avoid unnecessary prompt-side FHE work while preserving the ciphertext-only accelerator boundary during decode.

\Cref{fig:pd-serving-split} makes the workload consequence explicit. Prompt prefill is matrix-heavy and batched-GEMM dominated, whereas decode performs much smaller per-token compute while repeatedly touching cached state. Under \policyunified{}, every prompt token still pays the TEE-to-accelerator boundary cost and the encrypted linear path, so TTFT grows roughly with prompt length. Under \policypd{}, TTFT is closer to ``TEE prefill + one hybrid decode step,'' while steady-state decode remains governed by the same encrypted backbone.

Concretely, the offline phase pre-encodes model weights into GPU-resident BSGS-GEMM objects (approximately 48\,s for Qwen3 (0.6B)). Online, Phase~1 runs prompt prefill entirely on the CPU TEE at approximately 58\,ms/token, while Phase~2 invokes four FHE linear projections per layer per decode token at approximately 700\,ms per call.

\subsection{Threat Model and Leakage Contract}
\label{sec:threat}

\runhead{Trusted and untrusted components}
\policyunified{} trusts the attested CPU TEE, its firmware and attestation chain, and the in-TEE handling of FHE key material. We do not trust the host operating system, hypervisor, device runtime, GPU memory, PCIe-facing software stack, or the FHE accelerator. The accelerator may be a GPU, FPGA, or ASIC backend; in all cases, it is treated as an untrusted ciphertext evaluator rather than as a trusted execution environment. This is the sense in which \policyunified{} is a ``CPU TEE + FHE accelerator'' system: FHE protects the accelerator boundary, while the CPU TEE remains the root of trust.

\runhead{Protection goal}
The primary confidentiality goal is user-input privacy. \policyunified{} keeps prompts, intermediate activations, persistent cache state selected for the ciphertext path, and TEE-sealed cryptographic material hidden from the untrusted host and accelerator stack. Plaintext appears only inside the CPU TEE. The accelerator observes ciphertexts, deterministic FHE kernel launches, and coarse traffic metadata, but not plaintext activations or prompts.

\runhead{Model weights}
The default deployment assumes public model architecture and weights, as in many private-input inference settings. This matters because \policyunified{}'s FHE linear path uses ciphertext--plaintext multiplication with pre-encoded weight diagonals. Those diagonals are prepared inside the TEE, but the prototype should not be claimed to hide model weights from an accelerator that can inspect plaintext FHE operands. Private-model deployments can instead keep weight handling on the TEE path, use a trusted accelerator, or replace ciphertext--plaintext kernels with costlier ciphertext--ciphertext kernels. The main evaluation focuses on protecting user prompts and intermediate states.

\runhead{Persistent cache state}
The trusted-TEE / ciphertext-only accelerator boundary also covers persistent serving state. The \texttt{ShadowCipherKV} path is a simulated route used to validate serving-stack plumbing and boundary behavior; it is not a CKKS performance claim. Real CKKS KV and packed CKKS KV overheads are reported separately in the evaluation and appendix.

\runhead{Leakage contract}
The prototype hides prompt content and intermediate activations from the host and accelerator; plaintext appears only inside TEE windows or on the ciphertext path. It does not hide prompt/decode length, request size, ciphertext volume, coarse timing, or per-token kernel counts. Under the default public-weight setting, encoded plaintext FHE operands may also be accelerator-visible. Padding, bucketing, dummy kernels, ORAM-style padding, constant-time TEE routines, trusted accelerators, and ciphertext--ciphertext kernels are optional hardening mechanisms complementary to \policyunified{}'s operator-affinity and PD-split contributions.

\runhead{Out of scope}
As in most TEE-backed systems, we do not claim protection against physical attacks, denial-of-service, rollback or fork attacks, supply-chain compromise, microarchitectural side channels, or coarse request metadata such as length, traffic volume, and timing. Defending these channels requires padding, oblivious memory primitives, constant-time TEE implementations, or dummy-kernel emission, which are orthogonal to the trust-boundary partitioning studied here.

\runhead{Why DtE and PD do not widen the trust boundary}
The TEE already requires decryption authority to execute non-linear operators and attention-side control logic exactly. DtE therefore does not introduce a new trust root; it reuses the existing CPU TEE boundary to replace homomorphic refresh. Likewise, the PD split changes only when prompt-side state enters the hybrid decode path. It moves prefill into the same trusted TEE and leaves the accelerator-visible decode boundary unchanged.

\FloatBarrier

\section{Implementation}
\label{sec:impl}

\subsection{Software Stack}
\label{sec:impl-stack}

Our prototype is implemented atop nano-vLLM, a lightweight runtime that preserves the request/sequence handling, scheduler skeleton, and block-management substrate popularized by vLLM \cite{kwonEfficientMemoryManagement2023, yuGeeeekExplorerNanovllm2026}. The main extension is a \texttt{SecureLLM} engine that acts as a drop-in replacement for \texttt{nanovllm.LLM}. A \texttt{SecureInferenceConfig} object selects the secure policy, linear backend, KV mode, and run metadata. In the evaluation harness, the code-level policies \texttt{pure\_fhe}, \texttt{hybrid\_unified}, and \texttt{hybrid\_pd\_split} correspond directly to the paper-facing systems \policyfhe{}, \policyunified{}, and \policypd{}.

The runtime treats the KV cache as a first-class system component rather than as an external assumption. It exposes a pluggable \texttt{KVBackend} abstraction via \texttt{PlaintextCPUKV} and \texttt{ShadowCipherKV} backends. This is what lets \policyunified{} and \policypd{} share the same serving substrate across prefill, decode, and PD handoff: they do not merely assume a logical cache, but instantiate it explicitly in the runtime.

\subsection{FHE Backend}
\label{sec:impl-fhe}

The FHE backend builds on HEonGPU \cite{cryptoeprint:2024/1543}, an open-source GPU-accelerated CKKS library that provides the underlying homomorphic arithmetic, NTT kernels, and key-switching primitives used by both \policyunified{} and \policypd{}.

The prototype uses a CKKS parameter set with polynomial degree $N = 8192$, yielding 4096 complex slots, coefficient-modulus primes of $[60,30,30,30]$ bits, initial scale $\Delta = 2^{30}$, and one 60-bit special prime for key switching. This configuration targets an estimated security level of approximately 128\,bits under the RLWE hardness assumption used by the HE standard \cite{cheonHomomorphicEncryptionArithmetic2017}. On top of that substrate, the runtime is organized around a \emph{context pool} keyed by input dimension. For each unique projection shape, it provisions a CKKS context with the corresponding modulus chain and BSGS rotation parameters, then pre-encodes the weight matrix diagonals into GPU-resident BSGS-GEMM objects. For Qwen3 (0.6B) ($L{=}28$ layers, four projections each), this produces 112 persistent objects and takes approximately 48\,s on the H20 GPU at model-load time. Once encoded, these objects remain GPU-resident for all subsequent tokens, avoiding per-token re-encoding. Each per-operator FHE linear call then costs approximately 30\,ms for CPU-side encrypt, 500--800\,ms for the GPU BSGS matrix-vector multiply, and 30\,ms for CPU-side decrypt.

This organization matches the two dominant deployment regimes seen in evaluation. Fully cached models (e.g., GPT-2 (124M) with all 12 layers resident) behave like clean GEMM-bound cases with 3.1--4.7$\times$ speedup over per-token re-encoding. Partially cached models (e.g., Qwen3 (0.6B) with 23 of 28 layers cached) pay a measurable weight-encoding penalty for the remaining fallback layers. The implementation therefore exposes both persistent-cache and fallback-encode modes rather than assuming that all encrypted weights always reside on the device.

\subsection{TEE Integration}
\label{sec:impl-tee}

In the primary measurement configuration, the trusted side of the prototype runs under Intel TDX on the CPU. The TEE is responsible for attestation, FHE key release, policy control, boundary scheduling, and execution of all TEE-resident operators. Decryption and re-encryption buffers are managed inside the trusted runtime so that host-visible components only observe ciphertext movement and fixed control flow.

This integration model also explains why DtE is a systems feature rather than a paper-only abstraction. The TEE already needs access to plaintext activations for normalization, Softmax-style reductions, RoPE handling, and other control-heavy operators. Reusing that same authority for refresh keeps the trust boundary unchanged while avoiding the cost of homomorphic bootstrapping inside the accelerator path.

\subsection{API and Serving Path}
\label{sec:impl-api}

The serving interface remains OpenAI-compatible. A host-facing API server receives requests, forwards them into the trusted runtime, and returns generated tokens through the same request path used for ordinary serving. Inside the TEE, request handling, batching metadata, policy selection, and KV management share one control path; this is what allows \policyunified{} and \policypd{} to be presented as routing policies inside a common runtime rather than as separate model rewrites.

Phase-aware routing is implemented at the serving layer. Under \policyunified{}, prompt tokens and decode tokens both use the hybrid path. Under \policypd{}, prompt prefill stays in the TEE, prompt-side KV state is materialized there, and only the decode-side handoff enters the FHE path. The decode kernels themselves are unchanged.

\FloatBarrier

\section{Evaluation}
\label{sec:eval}

\subsection{Measurement Methodology and Baselines}
\label{sec:eval-setup}

\runhead{Measurement hierarchy}
The evaluation separates direct CKKS/FHE measurements, projected \policyfhe{} baselines, estimator-style comparison rows, and CPU endpoint rows. Direct CKKS/FHE measurements support the main \policyunified{} and \policypd{} results. \policyfhe{} rows are projections from measured primitive costs and model operator counts, rather than direct serving deployments. Estimator-style rows follow Euston's methodology for external comparison. CPU endpoint rows characterize the plaintext latency floor under the CPU-TEE trust assumption.

\runhead{Why direct models are smaller than estimator rows}
Direct deployment materializes FHE serving state, not just raw parameters: encoded BSGS diagonals, rotation material, workspace, persistent KV state, and fallback paths compete for accelerator memory. This is why Qwen3 (0.6B) already enters a partially cached regime and GPT-2 (774M) reaches the memory wall on a 96~GiB H20 after only a subset of layers are encoded.

\runhead{Hardware}
All primary measurements run on Alibaba Cloud using a single server with 24 vCPUs, 128~GiB system memory, and one NVIDIA H20 GPU with 96~GiB device memory. The CPU side runs under Intel TDX. The H20 instance is CC-capable as provisioned, but Bifrost's trusted computing base remains the CPU TEE. The GPU remains an untrusted FHE accelerator and sees only ciphertexts and accelerator-side metadata.

\runhead{Models}
Primary direct measurements use GPT-2 checkpoints \cite{gpt-2-radford2019language} and Qwen3 (0.6B) \cite{yangQwen3TechnicalReport2025}. For GPT-2 (124M) and Qwen3 (0.6B), \policyunified{} and \policypd{} are direct CKKS/FHE rows, while \policyfhe{} rows are projected. Qwen3 (8B) \cite{yangQwen3TechnicalReport2025} and Qwen3.5 (9B) \cite{qwen35blog} provide scale and CPU-endpoint anchors only; they are not main \policyunified{} or \policypd{} FHE result rows. GPT-2 (1.5B) plus LLaMA~3 (8B) appear only as estimator-style continuity rows for Euston comparison.

\runhead{Metrics and prompts}
We report time-to-first-token, total latency, and steady-state decode. The main body focuses on prompt lengths 16 and 64 because they expose the TTFT difference between unified hybrid execution and the PD-oriented split. Prompt-length-1 rows and additional scale/KV rows appear in the appendix.

\runhead{Construction of the projected \policyfhe{} baseline}
The \policyfhe{} rows are built from measured primitive latencies for homomorphic linear kernels, non-linear approximations, and refresh, combined with per-model operator counts and prompt/decode invocation structure. This preserves the Pure FHE execution semantics of the Euston-style methodology while avoiding the false impression that the \policyfhe{} rows are direct end-to-end measurements.

\runhead{KV cache in the serving stack}
\policyunified{} and \policypd{} do not treat the cache as a paper-only abstraction: the runtime materializes KV state through the same serving substrate that handles request routing and model execution. We report simulated \texttt{ShadowCipherKV} route checks separately from real CKKS encrypted-KV and packed-KV measurements. Packed CKKS KV on GPT-2 (124M) adds 0.248\% end-to-end overhead in API mode, while the Qwen3.5 (9B) CKKS KV measurement reports 60.6\,ms of KV overhead per step with all correctness checks passing.

\subsection{Cross-Model Latency Landscape}
\label{sec:cross-model}

\begin{figure}[htbp]
\centering
\begin{tikzpicture}
\begin{axis}[
    width=\linewidth,
    height=0.72\linewidth,
    ybar,
    bar width=8pt,
    ymode=log,
    ymin=10,
    ymax=1500,
    ylabel={CPU TEE decode latency (ms/token)},
    symbolic x coords={g2s,g2m,g2l,q06,q8,q9},
    xtick=data,
    xticklabels={GPT-2\\(124M),GPT-2\\(355M),GPT-2\\(774M),Qwen3\\(0.6B),Qwen3\\(8B),Qwen3.5\\(9B)},
    x tick label style={align=center, font=\scriptsize},
    ymajorgrids=true,
    grid style={gray!30},
    enlarge x limits=0.12,
    every axis plot/.append style={fill=TEEColor!80, draw=none},
]
\addplot coordinates {
    (g2s,13.74)
    (g2m,29.9)
    (g2l,59.2)
    (q06,58.55)
    (q8,490)
    (q9,429)
};

\node[font=\scriptsize, align=center, anchor=south] at (axis cs:q8,520) {466--490\\ms/tok};
\node[font=\scriptsize, align=center, anchor=south] at (axis cs:q9,460) {429\\ms/tok};
\draw[BoundaryGray, thick, <->] (axis cs:q8,1100) -- (axis cs:q9,1100);
\node[fill=white, font=\scriptsize, inner sep=1.5pt] at (axis cs:q8,1250) {9B is 8\% faster than 8B};
\end{axis}
\end{tikzpicture}
\caption{\textbf{Cross-model CPU endpoint decode landscape from 124M to $>$8B.} These rows characterize the CPU-TEE latency floor under the paper's trust assumption. They are not Bifrost targets to outperform.}
\Description{Cross-model log-scale bar chart of CPU TEE decode latency for GPT-2 (124M), GPT-2 (355M), GPT-2 (774M), Qwen3 (0.6B), Qwen3 (8B), and Qwen3.5 (9B), with an annotation for the Qwen3.5 (9B) versus Qwen3 (8B) comparison.}
\label{fig:cross-scale-landscape}
\end{figure}
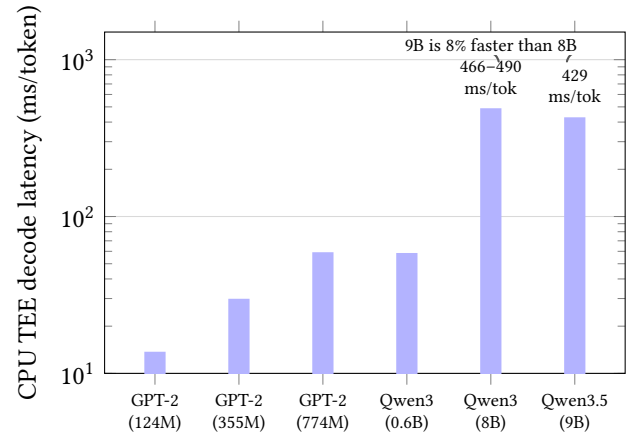

\vspace{-5pt}

\Cref{fig:cross-scale-landscape} establishes the trusted-CPU latency floor.
CPU endpoint decode scales from 13.74\,ms/tok on GPT-2 (124M) to
490\,ms/tok on Qwen3 (8B). These rows are included to show the plaintext floor;
encrypted Bifrost, Bifrost+, and Pure FHE rows are reported elsewhere with
separate measured/projected roles.
Qwen3.5 (9B) reaches 429\,ms/tok, 8\% faster than Qwen3 (8B) despite
a larger parameter count; a separate warm serving-path measurement
at 371--388\,ms/tok corroborates the benchmark figure.

\subsection{GEMM Tiling via TEE Scheduling}
\label{sec:tiling}

Homomorphic matrix multiplication dominates the runtime of transformer inference. We benchmark a representative CKKS GEMM workload of size $1024\times1024$ under different tiling granularities. The baseline performs one monolithic homomorphic GEMM. Tiling instead partitions the matrices into smaller tiles, executes tile-level homomorphic GEMMs on the GPU, and aggregates the outputs homomorphically.

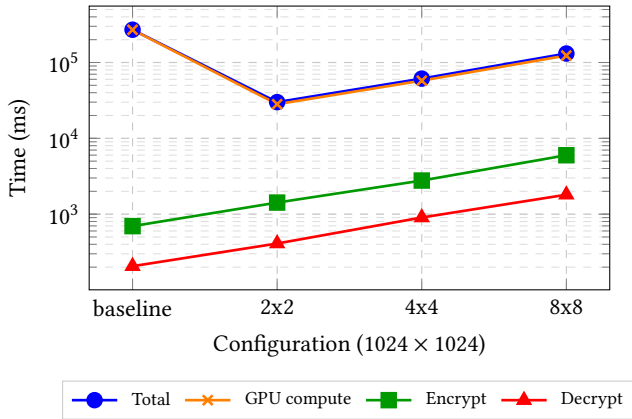
\begin{figure}[htbp]
\centering
\begin{tikzpicture}
\begin{axis}[
    width=\linewidth,
    height=0.63\linewidth,
    ymode=log,
    xlabel={Configuration ($1024\times1024$)},
    ylabel={Time (ms)},
    symbolic x coords={baseline,2x2,4x4,8x8},
    xtick=data,
    grid=both,
    major grid style={gray!45,dashed},
    minor grid style={gray!25,dashed},
    legend style={at={(0.5,-0.32)}, anchor=north, legend columns=4, font=\scriptsize, draw=gray!40, fill=white, column sep=3pt},
    tick label style={font=\small},
    label style={font=\small},
    every axis plot/.append style={line width=1pt, mark size=2.6pt},
]
\addplot[color=blue,mark=*] coordinates {
    (baseline,270572.000)
    (2x2,30021.021)
    (4x4,61304.664)
    (8x8,131302.547)
};
\addlegendentry{Total}

\addplot[color=orange,mark=x] coordinates {
    (baseline,269671.000)
    (2x2,28190.021)
    (4x4,57627.664)
    (8x8,123514.547)
};
\addlegendentry{GPU compute}

\addplot[color=green!60!black,mark=square*] coordinates {
    (baseline,695.000)
    (2x2,1421.000)
    (4x4,2774.000)
    (8x8,5983.000)
};
\addlegendentry{Encrypt}

\addplot[color=red,mark=triangle*] coordinates {
    (baseline,206.000)
    (2x2,410.000)
    (4x4,903.000)
    (8x8,1805.000)
};
\addlegendentry{Decrypt}
\end{axis}
\end{tikzpicture}
\caption{Runtime breakdown of a $1024\times1024$ homomorphic GEMM under different tiling granularities. The 2$\times$2 configuration is the best point in our setting, reducing GPU compute time from 269.7\,s to 28.2\,s and end-to-end time from 270.6\,s to 30.0\,s.}
\Description{Log-scale plot of homomorphic GEMM latency under different tiling granularities. The best point is the 2 by 2 tiling configuration.}
\label{fig:gemm_tiling}
\end{figure}

For this CKKS GEMM, 2$\times$2 tiling is the sweet spot: it improves GPU homomorphic compute efficiency enough to outweigh the extra encryption and decryption work, yielding a 9.01$\times$ speedup. More aggressive tiling (4$\times$4, 8$\times$8) leads to rapidly increasing tile-level operations that erode the benefit. This is precisely why the scheduler must choose tiling jointly with TEE-boundary cost rather than treating tile size as a pure accelerator-kernel decision.

\subsection{Non-Linear Operators and DtE Refresh}
\label{sec:nonlinear}

A key motivation for hybrid execution is that non-linear operators and ciphertext refresh are expensive, and sometimes numerically fragile, under pure FHE. Table~\ref{tab:nl-main} summarizes representative operator-level measurements using the GPU FHE path and the TEE path used by our hybrid system.

\begin{table}[htbp]
\caption{Representative operator and refresh microbenchmarks. These are primitive/operator provenance rows, not Bifrost end-to-end serving results. FHE rows use the GPU backend; TEE rows execute the exact plaintext operator or DtE refresh inside the trusted CPU boundary.}
\label{tab:nl-main}
\centering
\small
\begin{tabular}{lrrc}
\toprule
\textbf{Primitive} & \textbf{GPU FHE (s)} & \textbf{TEE (s)} & \textbf{Speedup} \\
\midrule
GELU & 0.02694 & 0.000068 & 398$\times$ \\
LayerNorm & 0.01559 & 0.000678 & 23$\times$ \\
Softmax & 0.00935 & 0.000537 & 17$\times$ \\
Refresh & 0.08701 & 0.00659 & 13.2$\times$ \\
\bottomrule
\end{tabular}
\end{table}

TEE offloading substantially reduces both runtime and numerical risk for non-linear operators. Compared to GPU-based FHE evaluation, TEE execution is about 398$\times$ faster for GELU, 23$\times$ faster for LayerNorm, and 17$\times$ faster for Softmax in our measured setting. For refresh, DtE is 13.2$\times$ faster than GPU bootstrapping. The latency advantage alone would already justify hybrid execution, but the qualitative effect is equally important: the TEE path executes the native operator exactly, avoiding the approximation drift that becomes especially problematic for operators such as Softmax.

\subsection{Comparison to Pure-FHE Baselines}
\label{sec:e2e-baseline}

\begin{table}[htbp]
\caption{Estimator-style continuity comparison against the \policyfhe{} SOTA baseline Euston. These rows are projected from the older methodology and serve only as an external baseline reference; they are not direct Bifrost end-to-end deployment results.}
\label{tab:euston-baseline}
\centering
\small
\begin{tabular}{lccc}
\toprule
\textbf{Model} & \textbf{\policyfhe{} (s)} & \textbf{\policyunified{} (s)} & \textbf{Speedup} \\
\midrule
GPT-2 (1.5B) & 555.544 & 60.048 & 9.252$\times$ \\
LLaMA~3 (8B) & 2804.887 & 283.157 & 9.906$\times$ \\
\bottomrule
\end{tabular}
\end{table}

\vspace{-5pt}

\begin{table*}[htbp]
\caption{Detailed stepwise end-to-end comparison for the main prompt lengths. Each cell reports TTFT / total latency / steady-state decode in seconds. \policyfhe{} is projected from measured primitive costs; \policyunified{} and \policypd{} are direct CKKS/FHE measurements.}
\label{tab:e2e-main}
\centering
\small
\begin{tabular}{llccc}
\toprule
\textbf{Model} & \textbf{Prompt} & \textbf{\policyfhe{}} & \textbf{\policyunified{}} & \textbf{\policypd{}} \\
\midrule
GPT-2 (124M) & 16 & 579.190 / 1090.240 / 34.070 & 62.199 / 120.579 / 3.892 & 4.269 / 63.399 / 3.942 \\
GPT-2 (124M) & 64 & 2214.550 / 2725.600 / 34.070 & 249.849 / 308.859 / 3.934 & 5.459 / 63.659 / 3.880 \\
Qwen3 (0.6B) & 16 & 1946.840 / 2748.480 / 114.520 & 352.302 / 506.631 / 22.047 & 23.023 / 177.261 / 22.034 \\
Qwen3 (0.6B) & 64 & 7443.800 / 8245.440 / 114.520 & 1419.378 / 1574.631 / 22.179 & 26.585 / 182.335 / 22.250 \\
\bottomrule
\end{tabular}
\end{table*}

\Cref{tab:euston-baseline} restores the external \policyfhe{} comparison in the same estimation setting used by Euston. Under that continuity setting, \policyunified{} achieves a consistent $\approx 9\times$ reduction in estimated end-to-end latency. We keep this comparison as projection context only; direct Bifrost claims use direct CKKS/FHE measurements instead of Euston operator or estimator rows.

\subsection{Real End-to-End Hybrid Deployment}
\label{sec:e2e}

\begin{figure}[htbp]
\centering
\begin{tikzpicture}
\begin{axis}[
    width=\linewidth,
    height=0.72\linewidth,
    ybar,
    bar width=7pt,
    ymode=log,
    ymin=10,
    ymax=10000,
    ylabel={Total latency (s)},
    symbolic x coords={g16,g64,q16,q64},
    xtick=data,
    xticklabels={GPT-2\\(124M)\\$P{=}16$,GPT-2\\(124M)\\$P{=}64$,Qwen3\\(0.6B)\\$P{=}16$,Qwen3\\(0.6B)\\$P{=}64$},
    x tick label style={align=center, font=\scriptsize},
    ymajorgrids=true,
    grid style={gray!30},
    legend style={at={(0.5,-0.25)}, anchor=north, legend columns=3, draw=gray!40, fill=white, font=\scriptsize, column sep=3pt},
    enlarge x limits=0.16,
]
\addplot[fill=FHEColor!80, draw=none] coordinates {
    (g16,1090.240)
    (g64,2725.600)
    (q16,2748.480)
    (q64,8245.440)
};
\addlegendentry{\policyfhe{} (proj.)}

\addplot[fill=UnifiedColor!85, draw=none] coordinates {
    (g16,120.579)
    (g64,308.859)
    (q16,506.631)
    (q64,1574.631)
};
\addlegendentry{\policyunified{} (meas.)}

\addplot[fill=PDColor!85, draw=none] coordinates {
    (g16,63.399)
    (g64,63.659)
    (q16,177.261)
    (q64,182.335)
};
\addlegendentry{\policypd{} (meas.)}
\end{axis}
\end{tikzpicture}
\caption{\textbf{Stepwise end-to-end comparison.} \policyfhe{} bars are projected; \policyunified{} and \policypd{} bars are direct CKKS/FHE measurements.}
\Description{Log-scale grouped bar chart comparing Pure FHE, Bifrost, and Bifrost plus total latency for GPT-2 (124M) and Qwen3 (0.6B) at prompt lengths 16 and 64.}
\label{fig:main-result}
\end{figure}
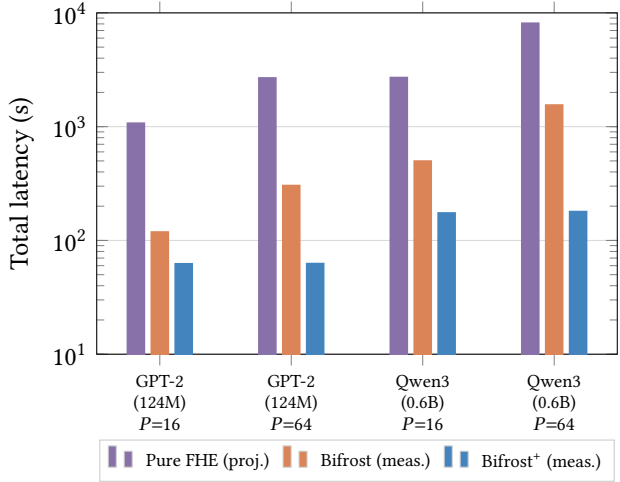

\Cref{fig:main-result} and \cref{tab:e2e-main} present the stepwise end-to-end result. Against the projected \policyfhe{} reference, \policyunified{} reduces total latency substantially (e.g., GPT-2 (124M) from 1090\,s to 121\,s at $P{=}16$) by removing non-linear operators and refresh from the homomorphic path. \policypd{} then further reduces it to 63\,s by moving prompt prefill into the TEE. The decode column confirms that neither hybrid policy changes the per-token decode cost---the gain comes from doing \emph{less} encrypted work, not from faster encrypted kernels.

\subsection{\texorpdfstring{\policypd{}}{Bifrost+} over \texorpdfstring{\policyunified{}}{Bifrost}}
\label{sec:pd}

\begin{table}[htbp]
\caption{\policypd{} TTFT measurements for direct CKKS/FHE PD-split rows. Unified TTFT is the same hybrid decode path without the prompt-side PD split. PD split removes prompt-side encrypted passes; it does not speed up the FHE kernels.}
\label{tab:pd-scaling}
\centering
\scriptsize
\begin{tabular}{lrrrr}
\toprule
\textbf{Model} & \textbf{Prompt} & \textbf{\policypd{} TTFT} & \textbf{Unified TTFT} & \textbf{Gain} \\
\midrule
GPT-2 (124M) & 1 & 3.888\,s & 7.769\,s & 2.0$\times$ \\
GPT-2 (124M) & 16 & 4.269\,s & 62.199\,s & 14.6$\times$ \\
GPT-2 (124M) & 64 & 5.459\,s & 249.849\,s & 45.8$\times$ \\
Qwen3 (0.6B) & 16 & 23.023\,s & 352.302\,s & 15.3$\times$ \\
Qwen3 (0.6B) & 64 & 26.585\,s & 1419.378\,s & 53.4$\times$ \\
\bottomrule
\end{tabular}
\end{table}

\runhead{Interpretation}
\Cref{tab:pd-scaling} shows that the gain grows with prompt length because \policypd{} avoids repeated prompt-side encrypted passes. This is a scheduling result rather than a kernel-speed result. \policyunified{} and \policypd{} use the same encrypted decode kernels, and steady-state decode changes by at most 2\%. The gain comes from deciding when \emph{not} to use the encrypted accelerator: prompt prefill stays in the CPU TEE, while only the decode-side handoff enters the hybrid path.

\section{Discussion}
\label{sec:discussion}

\subsection{Cost Structure and Memory Boundary}
\label{sec:cost}

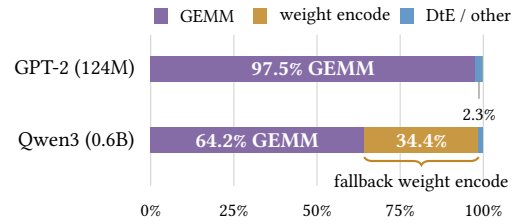
\begin{figure}[htbp]
\centering
\begin{tikzpicture}[x=0.44cm,y=1cm,font=\small]
    \foreach \x/\lbl in {0/0,2.5/25,5/50,7.5/75,10/100}{
        \draw[gray!35] (\x,0.4) -- (\x,2.4);
        \node[anchor=north, font=\scriptsize] at (\x,0.28) {\lbl\%};
    }

    \draw[fill=FHEColor!82, draw=none] (0.1,2.55) rectangle (0.5,2.75);
    \node[anchor=west, font=\scriptsize] at (0.6,2.65) {GEMM};
    \draw[fill=EncodeColor!85, draw=none] (3.1,2.55) rectangle (3.5,2.75);
    \node[anchor=west, font=\scriptsize] at (3.6,2.65) {weight encode};
    \draw[fill=PDColor!70, draw=none] (7.5,2.55) rectangle (7.9,2.75);
    \node[anchor=west, font=\scriptsize] at (8.0,2.65) {DtE / other};

    \node[anchor=east, font=\footnotesize] at (-0.2,1.95) {GPT-2 (124M)};
    \draw[fill=FHEColor!82, draw=none] (0,1.78) rectangle (9.75,2.12);
    \draw[fill=PDColor!70, draw=none] (9.75,1.78) rectangle (9.98,2.12);
    \draw[fill=TEEColor!65, draw=none] (9.98,1.78) rectangle (10,2.12);
    \node[white, font=\footnotesize\bfseries] at (4.88,1.95) {97.5\% GEMM};
    \draw[gray, thin] (9.87,1.78) -- (9.87,1.55);
    \node[anchor=north, font=\scriptsize] at (9.87,1.55) {2.3\%};

    \node[anchor=east, font=\footnotesize] at (-0.2,1.0) {Qwen3 (0.6B)};
    \draw[fill=FHEColor!82, draw=none] (0,0.83) rectangle (6.42,1.17);
    \draw[fill=EncodeColor!85, draw=none] (6.42,0.83) rectangle (9.86,1.17);
    \draw[fill=PDColor!70, draw=none] (9.86,0.83) rectangle (10,1.17);
    \node[white, font=\footnotesize\bfseries] at (3.2,1.0) {64.2\% GEMM};
    \node[white, font=\footnotesize\bfseries] at (8.14,1.0) {34.4\%};
    \draw[EncodeColor, thick, decorate, decoration={brace, amplitude=3pt, mirror}]
        (6.42,0.78) -- (9.86,0.78);
    \node[anchor=north, font=\scriptsize, align=center] at (8.14,0.68) {fallback weight encode};
\end{tikzpicture}
\caption{\textbf{Cost structure and systems boundary.} In fully cached GPT-2 (124M) FHE decode, GEMM dominates runtime and DtE is small. In Qwen3 (0.6B), five fallback layers cause a substantial weight-encoding penalty, exposing GPU memory capacity as a first-order systems constraint.}
\Description{Two stacked horizontal bars summarize decode costs: GPT-2 (124M) is dominated by GEMM, while Qwen3 (0.6B) includes a large fallback weight-encoding component due to incomplete GPU caching.}
\label{fig:cost}
\end{figure}

\begin{table}[htbp]
\caption{FHE decode cost breakdown. GPT-2 (124M) is fully cached; Qwen3 (0.6B) pays fallback weight encoding because not all encoded weights fit in accelerator-resident cache.}
\label{tab:fhe-cost-breakdown}
\centering
\scriptsize
\setlength{\tabcolsep}{2.5pt}
\begin{tabular}{lrrrr}
\toprule
\textbf{Model} & \textbf{Total} & \textbf{GEMM} & \textbf{DtE} & \textbf{Fallback} \\
\midrule
GPT-2 (124M), 12/12 & 3.88s & 3.78s (97.5\%) & 0.09s (2.3\%) & 0 \\
Qwen3 (0.6B), 23/28 & 22.01s & 14.14s (64.2\%) & 0.31s (1.4\%) & 7.56s (34.4\%) \\
\bottomrule
\end{tabular}
\end{table}

\Cref{tab:fhe-cost-breakdown} summarizes the FHE decode breakdown. In fully cached GPT-2 (124M) FHE decode, GEMM consumes 97.5\% of runtime and DtE only 2.3\%, so PD mainly removes repeated prompt-side GEMM. Qwen3 (0.6B) exposes the next constraint: with 23 of 28 layers cached, fallback weight encoding still contributes 34.4\% of runtime. This is encoded-weight/cache pressure, distinct from reference-route GPU memory utilization. Full accelerator-resident caching is therefore first-order. In the H20 96\,GiB budget, GPT-2 (774M) already reaches the memory wall: 23 of 36 layers occupy $\sim$92.1\,GiB after encoding. CPU TEE anchors remain faster, with Qwen3 (8B) at 466--490\,ms/token and Qwen3.5 (9B) at 429\,ms/token.

\subsection{Limitations}
\label{sec:limitations}

The prototype has four limitations: (1)~memory, not arithmetic, is the dominant boundary---once encoded weights and persistent state stop fitting on device, fallback encoding becomes a first-order cost; (2)~the evaluation is single-accelerator and single-tenant---multi-GPU partitioning and admission control remain future work; (3)~our threat model excludes rollback, denial-of-service, and microarchitectural side channels; (4)~encrypted KV overhead ($\sim$1.25\,MiB/token/layer for GPT-2 (124M)) is modest for short prompts but grows with longer contexts.

\section{Related Work}
\label{sec:related}

\runhead{FHE for neural network inference}
Private neural inference has progressed from interactive MPC/PHE systems such as MUSE \cite{lehmkuhlMuseSecureInference2021a} to non-interactive FHE systems. BOLT, NEXUS, BumbleBee, EncryptedLLM, and Euston improve communication, packing, and GPU utilization for secure transformer inference \cite{pangBOLTPrivacyPreservingAccurate2024, zhangSecureTransformerInference2025a, DBLP:conf/ndss/LuHGLLRHWC25, pmlr-v267-de-castro25a, gaoEustonEfficientUserFriendlyCorrected}. They provide the strongest pure-FHE baselines but remain constrained by non-linear evaluation, level management, and refresh cost.

\runhead{Multi-scheme and hybrid cryptographic approaches}
BLB combines MPC with CKKS-style HE to split transformer operators across schemes \cite{blb25usenixsecurity}. EncryptedLargeLanguage2025 pushes attention into server-side FHE while keeping the rest client-side \cite{EncryptedLargeLanguage2025}. TEEFHE and HT2ML show that TEEs can absorb bootstrapping or HE-unfriendly stages \cite{wangScalableFullyHomomorphic2019, wangHT2MLEfficientHybrid2023a}. TEE-only and CPU+GPU TEE systems are complementary endpoints for confidential LLM inference, and heterogeneous confidential-computing surveys highlight broader trust and attack-surface considerations for GPU TEEs \cite{chrapekConfidentialLLMInference2025,wangConfidentialComputingHeterogeneous2026}. TEE+MPC, masking, and trusted-accelerator designs occupy alternative points in the design space and may be better latency choices under different trust assumptions. Our narrower claim is that, to our knowledge, we are the first to instantiate TEE+FHE transformer/LLM inference with a CPU TEE plus accelerator-backed FHE path, and to study both operator-affinity and PD splitting while keeping the accelerator view ciphertext-only.

\runhead{FHE acceleration}
WarpDrive, Neo, and HEonGPU target GPU-accelerated CKKS; Trinity, Alchemist, FAB, and FAST explore domain-specific accelerators; and HEIR plus Orion reduce the compilation cost of encrypted execution \cite{fanWarpDriveGPUBasedFully2025a, jiaoNeoEfficientFully2025, cryptoeprint:2024/1543, dengTrinityGeneralPurpose2024, muAlchemistUnifiedAccelerator2024, agrawalFABFPGAbasedAccelerator2023, fanFASTFHEAccelerator2025, ali2025heir, ebelOrionFullyHomomorphic2025}. Building on HEonGPU, we ask which operators and phases should still pay encrypted execution cost once accelerator-backed encrypted linear algebra is practical.

\runhead{LLM serving}
vLLM demonstrates paged KV management for LLM deployment \cite{kwonEfficientMemoryManagement2023}, and DistServe/Splitwise establish PD disaggregation as a serving optimization \cite{zhongDistServeDisaggregatingPrefill2024, patelSplitwiseEfficientGenerative2024}. We adapt PD as a split across the TEE--FHE trust boundary.

\section{Conclusion}
\label{sec:conclusion}

This paper studies confidential Transformer/LLM serving where the CPU TEE is the only trusted execution boundary and the accelerator acts only as a ciphertext evaluator. Under this assumption, FHE is not a latency optimization over plaintext CPU TEE-only execution; it is the mechanism that lets an untrusted accelerator participate without seeing prompts or intermediate activations. \policyunified{} instantiates this design end to end by keeping linear layers on accelerator-backed FHE and moving non-linear operators and refresh into the CPU TEE. \policypd{} then applies PD splitting to keep prompt prefill inside the trusted CPU domain.

The systems lesson is selective encrypted execution: remove non-linear operators and refresh from the homomorphic critical path, then remove prompt-side encrypted passes from TTFT. In an estimator-style comparison matching Euston's methodology, this yields 9.25$\times$ and 9.91$\times$ projected latency reductions on GPT-2 (1.5B) and LLaMA~3 (8B). In direct deployments, \policypd{} reduces TTFT from 62.2\,s to 4.3\,s on GPT-2 (124M) and from 352.3\,s to 23.0\,s on Qwen3 (0.6B) at prompt length~16. The remaining boundary is memory, so future work should focus on encoded-weight residency, multi-accelerator partitioning, long-context KV management, and optional side-channel hardening.

\FloatBarrier

\clearpage
\appendix

\section{Generalization}
\label{sec:generalization}

The architecture is broader than the current GPU-backed prototype. Because the accelerator sees ciphertexts only, the same split applies to FPGA- and ASIC-based FHE backends as well as to future GPU generations. Likewise, the trusted-side logic is not tied specifically to Intel TDX: the same CPU-TEE role could be instantiated with AMD SEV-style deployments so long as attestation and key release support the same trust assumptions.

The operator-affinity argument should also generalize beyond the exact models in this paper. GPT-2, Qwen3, and Qwen3.5 already span classical transformer blocks, RMSNorm, RoPE, grouped-query attention, and mixed architectural motifs. Qwen3.5 (9B) is particularly informative as an SSM-heavy architecture-mix anchor \cite{qwen35blog}: SSM-style layers can remain entirely on the CPU TEE without FHE overhead, yielding a 3.7--4.4$\times$ FHE speedup over all-attention architectures of comparable size.

\FloatBarrier

\section{Algorithmic Details}
\label{sec:appendix-algorithms}

This section provides formal pseudocode for two system-level procedures that complement the boundary-aware scheduler presented in the main text (\cref{alg:boundary-aware-scheduler}).

\textbf{Operator partitioning.}
\Cref{alg:secure-partition} formalizes how each transformer operator is assigned to an execution domain. Non-linear operators that require plaintext access (normalization, positional encoding, attention, and activations) are routed to the CPU TEE, while the four linear projections per layer are routed to GPU FHE when the accelerator is untrusted. This partition is the basis of the operator-domain mapping summarized in \cref{tab:operator-domain} of the main text.

\begin{algorithm}[tbp]
\DontPrintSemicolon
\caption{Operator partitioning by execution domain}
\label{alg:secure-partition}
\KwIn{transformer model $M$ with $L$ layers, threat model $T$}
\KwOut{partition $P$ mapping each operator to a domain}
\ForEach{operator $\mathit{op}$ in $M$}{
  \uIf{$\mathit{op}.\mathit{type} \in \{\text{Norm, RoPE, Softmax, AttentionControl, Activation}\}$}{
    $P[\mathit{op}] \leftarrow \text{TEE\_PLAINTEXT}$\tcp*{must see plaintext}
  }
  \uElseIf{$\mathit{op}.\mathit{type} \in \{\text{Linear\_QKV, Linear\_O, Linear\_GateUp, Linear\_Down}\}$}{
    \lIf{$T.\mathit{gpu\_trusted}$}{$P[\mathit{op}] \leftarrow \text{GPU\_PLAIN}$}
    \lElse{$P[\mathit{op}] \leftarrow \text{GPU\_FHE}$}
  }
}
\Return{$P$}\;
\end{algorithm}

\textbf{PD-split serving protocol.}
\Cref{alg:pd-serve} describes the end-to-end \policypd{} serving protocol. The offline phase pre-encodes model weights into GPU-resident BSGS-GEMM objects once at model-load time. The online phase then runs in two stages: Phase~1 executes prompt prefill entirely on the CPU TEE using plaintext matmul (no FHE), while Phase~2 enters the hybrid decode loop where each layer invokes four FHE linear projections on the GPU and all remaining operators execute on the CPU TEE.

\begin{algorithm}[tbp]
\DontPrintSemicolon
\caption{Hybrid PD-split serving protocol}
\label{alg:pd-serve}
\KwIn{prompt $\mathit{tokens}[1..T]$, generation budget $n$, policy}
\KwOut{generated response tokens}
\tcp{Offline: encode weights once at model load}
$W \leftarrow \textsc{LoadModel}()$\;
$G, C \leftarrow \textsc{EncodeWeights}(W, \Theta)$\tcp*{GPU-persistent BSGS objects}
\BlankLine
\tcp{Phase 1: fast CPU TEE prefill (no FHE)}
\For{$t = 1$ \KwTo $T$}{
  $h \leftarrow \textsc{Block}(h, W, t)$\tcp*{CPU plaintext matmul}
  $\mathit{KV}[l] \leftarrow \textsc{Append}(k_l, v_l)$ for each layer $l$\;
}
$t_0 \leftarrow \textsc{Sample}(\textsc{LMHead}(h))$\;
\BlankLine
\tcp{Phase 2: FHE decode loop}
\For{$i = 1$ \KwTo $n - 1$}{
  \lIf{$t_{i{-}1} = \textsc{EOS}$}{\textbf{break}}
  \ForEach{layer $l$}{
    TEE ops: Norm, RoPE, Softmax/control, Gating\;
    GPU FHE: $\textsc{FHE\_Linear}(h, G[l], C)$ for each projection\;
  }
  $t_i \leftarrow \textsc{Sample}(\textsc{LMHead}(h))$\;
}
\Return{$[t_0, t_1, \ldots]$}\;
\end{algorithm}

\section{Supporting Evaluation Tables}
\label{sec:appendix-tables}

This section collects supplementary evaluation data that supports the main-text results but is not essential to the primary narrative.

\runhead{Prompt-length-1 four-way matrix}
\Cref{tab:prompt1} reports TTFT, total latency, and steady-state decode for a single-token prompt across all four execution policies. These rows complement the prompt-16 and prompt-64 results in the main text by showing the limiting case where prefill cost is minimal and the PD advantage is smallest.

\begin{table}[htbp]
\caption{Prompt-length-1 rows. Each cell: TTFT\,/\,total\,/\,decode (s).}
\label{tab:prompt1}
\centering
\scriptsize
\begin{tabular}{lcccc}
\toprule
\textbf{Model} & \textbf{\policycpu{}} & \textbf{\policyfhe{}} & \textbf{\policyunified{}} & \textbf{\policypd{}} \\
\midrule
GPT-2 (124M) & .05 / .33 / .02 & 68 / 579 / 34 & 7.8 / 66 / 3.9 & 3.9 / 62 / 3.9 \\
Qwen3 (0.6B) & .11 / .52 / .06 & 229 / 1031 / 115 & 45 / 202 / 22 & 22 / 177 / 22 \\
\bottomrule
\end{tabular}
\end{table}

\runhead{CPU TEE and large-model scaling references}
\Cref{tab:cpu-scale} reports CPU endpoint latency across model scales. These rows are included as the plaintext latency floor under the CPU-TEE trust assumption, not as Bifrost measurements. Notably, Qwen3.5 (9B) is 8\% faster than Qwen3 (8B) despite its larger parameter count, likely because its SSM-heavy architecture mix \cite{qwen35blog} reduces the attention-heavy workload that dominates at this scale.

\begin{table}[htbp]
\caption{CPU endpoint latency-floor reference across scales. Bifrost does not try to outperform this endpoint; it targets ciphertext-only accelerator delegation.}
\label{tab:cpu-scale}
\centering
\footnotesize
\begin{tabularx}{\columnwidth}{@{}p{0.26\columnwidth}p{0.22\columnwidth}X@{}}
\toprule
\textbf{Model} & \textbf{CPU endpoint} & \textbf{Role} \\
\midrule
GPT-2 (124M) & 13.74\,ms/tok & primary CPU baseline JSON \\
Qwen3 (0.6B) & 58.55\,ms/tok & primary CPU baseline JSON, BF16 row \\
Qwen3 (8B) & 466--490\,ms/tok & scale/reference CPU endpoint; not a Bifrost/FHE main result \\
Qwen3.5 (9B) & 429\,ms/tok & scale/reference CPU endpoint; not a Bifrost/FHE main result \\
\bottomrule
\end{tabularx}
\end{table}

\runhead{KV measurement separation}
\Cref{tab:kv-evidence} separates the three KV measurement types used in the paper. Simulated \texttt{ShadowCipherKV} rows exercise route plumbing and boundary behavior only; real CKKS KV rows support encrypted-cache overhead and correctness claims; packed CKKS KV rows support packing and memory-reduction claims.

\begin{table}[htbp]
\caption{KV-cache measurement classes. Simulated \texttt{ShadowCipherKV}, real CKKS encrypted KV, and packed CKKS KV are separate classes and should not be collapsed into one performance claim.}
\label{tab:kv-evidence}
\centering
\scriptsize
\begin{tabularx}{\columnwidth}{@{}p{0.34\columnwidth}X@{}}
\toprule
\textbf{KV measurement} & \textbf{Metric and role} \\
\midrule
ShadowCipherKV route & Simulated infrastructure for route plumbing and boundary behavior only; no CKKS performance claim. \\
GPT-2 (124M) CKKS KV & Real CKKS encrypted KV microbenchmark; append 2.07\,ms, read 4.74\,ms, correctness passing. \\
GPT-2 (124M) packed KV & Packed real CKKS KV; 0.248\% overhead, 5.32$\times$ memory reduction, and 16/16 token agreement. \\
Qwen3.5 (9B) CKKS KV & Real CKKS encrypted KV; 60.6\,ms/step, 0.046\% overhead, and 32/32 token agreement. \\
\bottomrule
\end{tabularx}
\end{table}

\runhead{Ciphertext refresh microbenchmark}
\Cref{tab:bootstrapping-perf} compares three refresh strategies: GPU-based bootstrapping, CPU-based bootstrapping, and TEE-assisted DtE. DtE is 13.2$\times$ faster than GPU bootstrapping and over 4\,300$\times$ faster than CPU bootstrapping, which is the primary reason the hybrid architecture replaces homomorphic refresh with DtE.

\begin{table}[htbp]
\caption{Ciphertext refresh microbenchmark.}
\label{tab:bootstrapping-perf}
\centering
\footnotesize
\begin{tabular}{llr}
\toprule
\textbf{Primitive} & \textbf{Backend} & \textbf{Avg. time (s)} \\
\midrule
Bootstrapping & GPU FHE & 0.0870 \\
Bootstrapping & CPU FHE & 28.623 \\
DtE refresh & CPU TEE & 0.00659 \\
\bottomrule
\end{tabular}
\end{table}

\runhead{Non-linear operator latencies}
The final table reports per-operator latency for GELU, LayerNorm, and Softmax across CPU FHE, GPU FHE, and TEE plaintext backends. TEE execution is orders of magnitude faster than either FHE backend and preserves native operator semantics exactly, whereas FHE approximations introduce numerical deviation that grows especially large for Softmax. This is why the main body treats exact TEE execution of non-linear operators as both a latency optimization and a numerical-stability advantage.

\begin{table}[htbp]
\caption{Representative non-linear operator latencies across backends.}
\label{tab:nonlinear-perf}
\centering
\footnotesize
\begin{tabular}{llr}
\toprule
\textbf{Operator} & \textbf{Backend} & \textbf{Avg. time (s)} \\
\midrule
GELU & FHE (CPU) & 2.766 \\
GELU & FHE (GPU) & 0.0269 \\
GELU & TEE plaintext & $6.77\times 10^{-5}$ \\
\addlinespace
LayerNorm & FHE (CPU) & 2.241 \\
LayerNorm & FHE (GPU) & 0.0156 \\
LayerNorm & TEE plaintext & $6.78\times 10^{-4}$ \\
\addlinespace
Softmax & FHE (CPU) & 1.090 \\
Softmax & FHE (GPU) & 0.00935 \\
Softmax & TEE plaintext & $5.37\times 10^{-4}$ \\
\bottomrule
\end{tabular}
\end{table}


\begin{thebibliography}{32}

\ifx \showCODEN    \undefined \def \showCODEN     #1{\unskip}     \fi
\ifx \showISBNx    \undefined \def \showISBNx     #1{\unskip}     \fi
\ifx \showISBNxiii \undefined \def \showISBNxiii  #1{\unskip}     \fi
\ifx \showISSN     \undefined \def \showISSN      #1{\unskip}     \fi
\ifx \showLCCN     \undefined \def \showLCCN      #1{\unskip}     \fi
\ifx \shownote     \undefined \def \shownote      #1{#1}          \fi
\ifx \showarticletitle \undefined \def \showarticletitle #1{#1}   \fi
\ifx \showURL      \undefined \def \showURL       {\relax}        \fi
\providecommand\bibfield[2]{#2}
\providecommand\bibinfo[2]{#2}
\providecommand\natexlab[1]{#1}
\providecommand\showeprint[2][]{arXiv:#2}

\bibitem[AMD({[n.\,d.]})]%
        {AMDSecureEncrypted}
 \bibinfo{year}{[n.\,d.]}\natexlab{}.
\newblock \bibinfo{title}{{{AMD Secure Encrypted Virtualization}} ({{SEV}})}.
\newblock
\urldef\tempurl%
\url{https://www.amd.com/en/developer/sev.html}
\showURL{%
\tempurl}


\bibitem[Int({[n.\,d.]})]%
        {IntelTrustDomaina}
 \bibinfo{year}{[n.\,d.]}\natexlab{}.
\newblock \bibinfo{title}{{{Intel}}\textregistered{} {{Trust Domain
  Extensions}} ({{Intel}}\textregistered{} {{TDX}})}.
\newblock
\urldef\tempurl%
\url{https://www.intel.com/content/www/us/en/developer/tools/trust-domain-extensions/overview.html}
\showURL{%
\tempurl}


\bibitem[Agrawal et~al\mbox{.}(2023)]%
        {agrawalFABFPGAbasedAccelerator2023}
\bibfield{author}{\bibinfo{person}{Rashmi Agrawal}, \bibinfo{person}{Leo {de
  Castro}}, \bibinfo{person}{Guowei Yang}, \bibinfo{person}{Chiraag Juvekar},
  \bibinfo{person}{Rabia Yazicigil}, \bibinfo{person}{Anantha Chandrakasan},
  \bibinfo{person}{Vinod Vaikuntanathan}, {and} \bibinfo{person}{Ajay Joshi}.}
  \bibinfo{year}{2023}\natexlab{}.
\newblock \showarticletitle{{{FAB}}: {{An FPGA-based Accelerator}} for
  {{Bootstrappable Fully Homomorphic Encryption}}}. In
  \bibinfo{booktitle}{\emph{2023 {{IEEE International Symposium}} on
  {{High-Performance Computer Architecture}} ({{HPCA}})}}.
  \bibinfo{pages}{882--895}.
\newblock
\showISSN{2378-203X}
\href{https://doi.org/10.1109/HPCA56546.2023.10070953}{doi:\nolinkurl{10.1109/HPCA56546.2023.10070953}}


\bibitem[Ali et~al\mbox{.}(2025)]%
        {ali2025heir}
\bibfield{author}{\bibinfo{person}{Asra Ali}, \bibinfo{person}{Jaeho Choi},
  \bibinfo{person}{Bryant Gipson}, \bibinfo{person}{Shruthi Gorantala},
  \bibinfo{person}{Jeremy Kun}, \bibinfo{person}{Wouter Legiest},
  \bibinfo{person}{Lawrence Lim}, \bibinfo{person}{Alexander Viand},
  \bibinfo{person}{Meron~Zerihun Demissie}, {and} \bibinfo{person}{Hongren
  Zheng}.} \bibinfo{year}{2025}\natexlab{}.
\newblock \bibinfo{title}{HEIR: A Universal Compiler for Homomorphic
  Encryption}.
\newblock
\showeprint[arxiv]{2508.11095}~[cs.CR]
\urldef\tempurl%
\url{https://arxiv.org/abs/2508.11095}
\showURL{%
\tempurl}


\bibitem[Bredehoft and Frery(2025)]%
        {EncryptedLargeLanguage2025}
\bibfield{author}{\bibinfo{person}{Roman Bredehoft} {and}
  \bibinfo{person}{Jordan Frery}.} \bibinfo{year}{2025}\natexlab{}.
\newblock \bibinfo{title}{Towards {{Encrypted Large Language Models}} with
  {{FHE}}}.
\newblock \bibinfo{howpublished}{https://huggingface.co/blog/encrypted-llm}.
\newblock


\bibitem[Cheon et~al\mbox{.}(2017)]%
        {cheonHomomorphicEncryptionArithmetic2017}
\bibfield{author}{\bibinfo{person}{Jung~Hee Cheon}, \bibinfo{person}{Andrey
  Kim}, \bibinfo{person}{Miran Kim}, {and} \bibinfo{person}{Yongsoo Song}.}
  \bibinfo{year}{2017}\natexlab{}.
\newblock \showarticletitle{Homomorphic {{Encryption}} for {{Arithmetic}} of
  {{Approximate Numbers}}}. In \bibinfo{booktitle}{\emph{Advances in
  {{Cryptology}} -- {{ASIACRYPT}} 2017}},
  \bibfield{editor}{\bibinfo{person}{Tsuyoshi Takagi} {and}
  \bibinfo{person}{Thomas Peyrin}} (Eds.). \bibinfo{publisher}{Springer
  International Publishing}, \bibinfo{address}{Cham},
  \bibinfo{pages}{409--437}.
\newblock
\showISBNx{978-3-319-70694-8}
\href{https://doi.org/10.1007/978-3-319-70694-8_15}{doi:\nolinkurl{10.1007/978-3-319-70694-8_15}}


\bibitem[Chrapek et~al\mbox{.}(2025)]%
        {chrapekConfidentialLLMInference2025}
\bibfield{author}{\bibinfo{person}{Marcin Chrapek}, \bibinfo{person}{Marcin
  Copik}, \bibinfo{person}{Etienne Mettaz}, {and} \bibinfo{person}{Torsten
  Hoefler}.} \bibinfo{year}{2025}\natexlab{}.
\newblock \showarticletitle{Confidential {{LLM Inference}}: {{Performance}} and
  {{Cost Across CPU}} and {{GPU TEEs}}}. In \bibinfo{booktitle}{\emph{2025
  {{IEEE International Symposium}} on {{Workload Characterization}}
  ({{IISWC}})}}. \bibinfo{pages}{84--98}.
\newblock
\showISSN{2835-2238}
\href{https://doi.org/10.1109/IISWC66894.2025.00017}{doi:\nolinkurl{10.1109/IISWC66894.2025.00017}}


\bibitem[\c{S}ah \"Ozcan and Sava\c{s}(2024)]%
        {cryptoeprint:2024/1543}
\bibfield{author}{\bibinfo{person}{Ali \c{S}ah \"Ozcan} {and}
  \bibinfo{person}{Erkay Sava\c{s}}.} \bibinfo{year}{2024}\natexlab{}.
\newblock \bibinfo{title}{{HEonGPU}: a {GPU}-based Fully Homomorphic Encryption
  Library 1.0}.
\newblock \bibinfo{howpublished}{Cryptology {ePrint} Archive, Paper 2024/1543}.
\newblock
\urldef\tempurl%
\url{https://eprint.iacr.org/2024/1543}
\showURL{%
\tempurl}


\bibitem[De~Castro et~al\mbox{.}(2025)]%
        {pmlr-v267-de-castro25a}
\bibfield{author}{\bibinfo{person}{Leo De~Castro}, \bibinfo{person}{Daniel
  Escudero}, \bibinfo{person}{Adya Agrawal}, \bibinfo{person}{Antigoni
  Polychroniadou}, {and} \bibinfo{person}{Manuela Veloso}.}
  \bibinfo{year}{2025}\natexlab{}.
\newblock \showarticletitle{{E}ncrypted{LLM}: Privacy-Preserving Large Language
  Model Inference via {GPU}-Accelerated Fully Homomorphic Encryption}. In
  \bibinfo{booktitle}{\emph{Proceedings of the 42nd International Conference on
  Machine Learning}} \emph{(\bibinfo{series}{Proceedings of Machine Learning
  Research}, Vol.~\bibinfo{volume}{267})},
  \bibfield{editor}{\bibinfo{person}{Aarti Singh}, \bibinfo{person}{Maryam
  Fazel}, \bibinfo{person}{Daniel Hsu}, \bibinfo{person}{Simon Lacoste-Julien},
  \bibinfo{person}{Felix Berkenkamp}, \bibinfo{person}{Tegan Maharaj},
  \bibinfo{person}{Kiri Wagstaff}, {and} \bibinfo{person}{Jerry Zhu}} (Eds.).
  \bibinfo{publisher}{PMLR}, \bibinfo{pages}{12677--12688}.
\newblock


\bibitem[Deng et~al\mbox{.}(2024)]%
        {dengTrinityGeneralPurpose2024}
\bibfield{author}{\bibinfo{person}{Xianglong Deng}, \bibinfo{person}{Shengyu
  Fan}, \bibinfo{person}{Zhicheng Hu}, \bibinfo{person}{Zhuoyu Tian},
  \bibinfo{person}{Zihao Yang}, \bibinfo{person}{Jiangrui Yu},
  \bibinfo{person}{Dingyuan Cao}, \bibinfo{person}{Dan Meng},
  \bibinfo{person}{Rui Hou}, \bibinfo{person}{Meng Li}, \bibinfo{person}{Qian
  Lou}, {and} \bibinfo{person}{Mingzhe Zhang}.}
  \bibinfo{year}{2024}\natexlab{}.
\newblock \showarticletitle{Trinity: A General Purpose FHE Accelerator}. In
  \bibinfo{booktitle}{\emph{2024 57th IEEE/ACM International Symposium on
  Microarchitecture (MICRO)}}. \bibinfo{pages}{338--351}.
\newblock
\href{https://doi.org/10.1109/MICRO61859.2024.00033}{doi:\nolinkurl{10.1109/MICRO61859.2024.00033}}


\bibitem[Ebel et~al\mbox{.}(2025)]%
        {ebelOrionFullyHomomorphic2025}
\bibfield{author}{\bibinfo{person}{Austin Ebel}, \bibinfo{person}{Karthik
  Garimella}, {and} \bibinfo{person}{Brandon Reagen}.}
  \bibinfo{year}{2025}\natexlab{}.
\newblock \showarticletitle{Orion: {{A Fully Homomorphic Encryption Framework}}
  for {{Deep Learning}}}. In \bibinfo{booktitle}{\emph{Proceedings of the 30th
  {{ACM International Conference}} on {{Architectural Support}} for
  {{Programming Languages}} and {{Operating Systems}}, {{Volume}} 2}}
  \emph{(\bibinfo{series}{{{ASPLOS}} '25})}. \bibinfo{publisher}{Association
  for Computing Machinery}, \bibinfo{address}{New York, NY, USA},
  \bibinfo{pages}{734--749}.
\newblock
\showISBNx{979-8-4007-1079-7}
\href{https://doi.org/10.1145/3676641.3716008}{doi:\nolinkurl{10.1145/3676641.3716008}}


\bibitem[Fan et~al\mbox{.}(2025b)]%
        {fanWarpDriveGPUBasedFully2025a}
\bibfield{author}{\bibinfo{person}{Guang Fan}, \bibinfo{person}{Mingzhe Zhang},
  \bibinfo{person}{Fangyu Zheng}, \bibinfo{person}{Shengyu Fan},
  \bibinfo{person}{Tian Zhou}, \bibinfo{person}{Xianglong Deng},
  \bibinfo{person}{Wenxu Tang}, \bibinfo{person}{Liang Kong},
  \bibinfo{person}{Yixuan Song}, {and} \bibinfo{person}{Shoumeng Yan}.}
  \bibinfo{year}{2025}\natexlab{b}.
\newblock \showarticletitle{{{WarpDrive}}: {{GPU-Based Fully Homomorphic
  Encryption Acceleration Leveraging Tensor}} and {{CUDA Cores}}}. In
  \bibinfo{booktitle}{\emph{2025 {{IEEE International Symposium}} on {{High
  Performance Computer Architecture}} ({{HPCA}})}}.
  \bibinfo{pages}{1187--1200}.
\newblock
\showISSN{2378-203X}
\href{https://doi.org/10.1109/HPCA61900.2025.00091}{doi:\nolinkurl{10.1109/HPCA61900.2025.00091}}


\bibitem[Fan et~al\mbox{.}(2025a)]%
        {fanFASTFHEAccelerator2025}
\bibfield{author}{\bibinfo{person}{Shengyu Fan}, \bibinfo{person}{Xianglong
  Deng}, \bibinfo{person}{Liang Kong}, \bibinfo{person}{Guiming Shi},
  \bibinfo{person}{Guang Fan}, \bibinfo{person}{Dan Meng}, \bibinfo{person}{Rui
  Hou}, {and} \bibinfo{person}{Mingzhe Zhang}.}
  \bibinfo{year}{2025}\natexlab{a}.
\newblock \showarticletitle{{{FAST}}:{{An FHE Accelerator}} for
  {{Scalable-parallelism}} with {{Tunable-bit}}}. In
  \bibinfo{booktitle}{\emph{Proceedings of the 52nd {{Annual International
  Symposium}} on {{Computer Architecture}}}} \emph{(\bibinfo{series}{{{ISCA}}
  '25})}. \bibinfo{publisher}{Association for Computing Machinery},
  \bibinfo{address}{New York, NY, USA}, \bibinfo{pages}{92--106}.
\newblock
\showISBNx{979-8-4007-1261-6}
\href{https://doi.org/10.1145/3695053.3731407}{doi:\nolinkurl{10.1145/3695053.3731407}}


\bibitem[Gao et~al\mbox{.}(2026)]%
        {gaoEustonEfficientUserFriendlyCorrected}
\bibfield{author}{\bibinfo{person}{Xinwen Gao}, \bibinfo{person}{Shaojing Fu},
  \bibinfo{person}{Lin Liu}, \bibinfo{person}{Zhuotao Liu},
  \bibinfo{person}{Yuchuan Luo}, {and} \bibinfo{person}{Yongjun Wang}.}
  \bibinfo{year}{2026}\natexlab{}.
\newblock \showarticletitle{{Euston: Efficient and User-Friendly Secure
  Transformer Inference with Non-Interactivity}}. In
  \bibinfo{booktitle}{\emph{2026 IEEE Symposium on Security and Privacy (SP)}}.
  \bibinfo{publisher}{IEEE Computer Society}, \bibinfo{pages}{899--918}.
\newblock
\href{https://doi.org/10.1109/SP63933.2026.00048}{doi:\nolinkurl{10.1109/SP63933.2026.00048}}


\bibitem[Gentry(2009)]%
        {gentryFullyHomomorphicEncryption2009a}
\bibfield{author}{\bibinfo{person}{Craig Gentry}.}
  \bibinfo{year}{2009}\natexlab{}.
\newblock \showarticletitle{Fully Homomorphic Encryption Using Ideal Lattices}.
  In \bibinfo{booktitle}{\emph{Proceedings of the Forty-First Annual {{ACM}}
  Symposium on {{Theory}} of Computing}} \emph{(\bibinfo{series}{{{STOC}}
  '09})}. \bibinfo{publisher}{Association for Computing Machinery},
  \bibinfo{address}{New York, NY, USA}, \bibinfo{pages}{169--178}.
\newblock
\showISBNx{978-1-60558-506-2}
\href{https://doi.org/10.1145/1536414.1536440}{doi:\nolinkurl{10.1145/1536414.1536440}}


\bibitem[Jiao et~al\mbox{.}(2025)]%
        {jiaoNeoEfficientFully2025}
\bibfield{author}{\bibinfo{person}{Dian Jiao}, \bibinfo{person}{Xianglong
  Deng}, \bibinfo{person}{Zhiwei Wang}, \bibinfo{person}{Shengyu Fan},
  \bibinfo{person}{Yi Chen}, \bibinfo{person}{Dan Meng}, \bibinfo{person}{Rui
  Hou}, {and} \bibinfo{person}{Mingzhe Zhang}.}
  \bibinfo{year}{2025}\natexlab{}.
\newblock \showarticletitle{Neo: {{Towards Efficient Fully Homomorphic
  Encryption Acceleration}} Using {{Tensor Core}}}. In
  \bibinfo{booktitle}{\emph{Proceedings of the 52nd {{Annual International
  Symposium}} on {{Computer Architecture}}}} \emph{(\bibinfo{series}{{{ISCA}}
  '25})}. \bibinfo{publisher}{Association for Computing Machinery},
  \bibinfo{address}{New York, NY, USA}, \bibinfo{pages}{107--121}.
\newblock
\showISBNx{979-8-4007-1261-6}
\href{https://doi.org/10.1145/3695053.3731408}{doi:\nolinkurl{10.1145/3695053.3731408}}


\bibitem[Kwon et~al\mbox{.}(2023)]%
        {kwonEfficientMemoryManagement2023}
\bibfield{author}{\bibinfo{person}{Woosuk Kwon}, \bibinfo{person}{Zhuohan Li},
  \bibinfo{person}{Siyuan Zhuang}, \bibinfo{person}{Ying Sheng},
  \bibinfo{person}{Lianmin Zheng}, \bibinfo{person}{Cody~Hao Yu},
  \bibinfo{person}{Joseph~E. Gonzalez}, \bibinfo{person}{Hao Zhang}, {and}
  \bibinfo{person}{Ion Stoica}.} \bibinfo{year}{2023}\natexlab{}.
\newblock \showarticletitle{Efficient Memory Management for Large Language
  Model Serving with {PagedAttention}}. In
  \bibinfo{booktitle}{\emph{Proceedings of the 29th Symposium on Operating
  Systems Principles}} \emph{(\bibinfo{series}{SOSP '23})}.
  \bibinfo{publisher}{ACM}, \bibinfo{pages}{611--626}.
\newblock
\href{https://doi.org/10.1145/3600006.3613165}{doi:\nolinkurl{10.1145/3600006.3613165}}


\bibitem[Lehmkuhl et~al\mbox{.}(2021)]%
        {lehmkuhlMuseSecureInference2021a}
\bibfield{author}{\bibinfo{person}{Ryan Lehmkuhl}, \bibinfo{person}{Pratyush
  Mishra}, \bibinfo{person}{Akshayaram Srinivasan}, {and}
  \bibinfo{person}{Raluca~Ada Popa}.} \bibinfo{year}{2021}\natexlab{}.
\newblock \showarticletitle{Muse: {{Secure Inference Resilient}} to {{Malicious
  Clients}}}. In \bibinfo{booktitle}{\emph{30th {{USENIX Security Symposium}}
  ({{USENIX Security}} 21)}}. \bibinfo{pages}{2201--2218}.
\newblock
\showISBNx{978-1-939133-24-3}


\bibitem[Lu et~al\mbox{.}(2025)]%
        {DBLP:conf/ndss/LuHGLLRHWC25}
\bibfield{author}{\bibinfo{person}{Wen{-}jie Lu}, \bibinfo{person}{Zhicong
  Huang}, \bibinfo{person}{Zhen Gu}, \bibinfo{person}{Jingyu Li},
  \bibinfo{person}{Jian Liu}, \bibinfo{person}{Cheng Hong},
  \bibinfo{person}{Kui Ren}, \bibinfo{person}{Tao Wei}, {and}
  \bibinfo{person}{Wenguang Chen}.} \bibinfo{year}{2025}\natexlab{}.
\newblock \showarticletitle{{BumbleBee: Secure Two-party Inference Framework
  for Large Transformers}}. In \bibinfo{booktitle}{\emph{32nd Annual Network
  and Distributed System Security Symposium, {NDSS} 2025}}.
  \bibinfo{publisher}{The Internet Society}.
\newblock


\bibitem[Mu et~al\mbox{.}(2024)]%
        {muAlchemistUnifiedAccelerator2024}
\bibfield{author}{\bibinfo{person}{Jianan Mu}, \bibinfo{person}{Husheng Han},
  \bibinfo{person}{Shangyi Shi}, \bibinfo{person}{Jing Ye},
  \bibinfo{person}{Zizhen Liu}, \bibinfo{person}{Shengwen Liang},
  \bibinfo{person}{Meng Li}, \bibinfo{person}{Mingzhe Zhang},
  \bibinfo{person}{Song Bian}, \bibinfo{person}{Xing Hu},
  \bibinfo{person}{Huaiwei Li}, {and} \bibinfo{person}{Xiaowei Li}.}
  \bibinfo{year}{2024}\natexlab{}.
\newblock \showarticletitle{Alchemist: {{A Unified Accelerator Architecture}}
  for {{Cross-Scheme Fully Homomorphic Encryption}}}. In
  \bibinfo{booktitle}{\emph{Proceedings of the 61st {{ACM}}/{{IEEE Design
  Automation Conference}}}} \emph{(\bibinfo{series}{{{DAC}} '24})}.
  \bibinfo{publisher}{Association for Computing Machinery},
  \bibinfo{address}{New York, NY, USA}, \bibinfo{pages}{1--6}.
\newblock
\showISBNx{979-8-4007-0601-1}
\href{https://doi.org/10.1145/3649329.3657331}{doi:\nolinkurl{10.1145/3649329.3657331}}


\bibitem[Pang et~al\mbox{.}(2024)]%
        {pangBOLTPrivacyPreservingAccurate2024}
\bibfield{author}{\bibinfo{person}{Qi Pang}, \bibinfo{person}{Jinhao Zhu},
  \bibinfo{person}{Helen M{\"o}llering}, \bibinfo{person}{Wenting Zheng}, {and}
  \bibinfo{person}{Thomas Schneider}.} \bibinfo{year}{2024}\natexlab{}.
\newblock \showarticletitle{{{BOLT}}: {{Privacy-Preserving}}, {{Accurate}} and
  {{Efficient Inference}} for {{Transformers}}}. In
  \bibinfo{booktitle}{\emph{2024 {{IEEE Symposium}} on {{Security}} and
  {{Privacy}} ({{SP}})}}. \bibinfo{pages}{4753--4771}.
\newblock
\showISSN{2375-1207}
\href{https://doi.org/10.1109/SP54263.2024.00130}{doi:\nolinkurl{10.1109/SP54263.2024.00130}}


\bibitem[Patel et~al\mbox{.}(2024)]%
        {patelSplitwiseEfficientGenerative2024}
\bibfield{author}{\bibinfo{person}{Pratyush Patel}, \bibinfo{person}{Esha
  Choukse}, \bibinfo{person}{Chaojie Zhang}, \bibinfo{person}{Aashaka Shah},
  \bibinfo{person}{\'I\~nigo Goiri}, \bibinfo{person}{Saeed Maleki}, {and}
  \bibinfo{person}{Ricardo Bianchini}.} \bibinfo{year}{2024}\natexlab{}.
\newblock \showarticletitle{Splitwise: Efficient Generative {LLM} Inference
  Using Phase Splitting}. In \bibinfo{booktitle}{\emph{Proceedings of the 51st
  Annual International Symposium on Computer Architecture}}
  \emph{(\bibinfo{series}{ISCA '24})}. \bibinfo{publisher}{ACM},
  \bibinfo{pages}{118--132}.
\newblock
\href{https://doi.org/10.1145/3620665.3640401}{doi:\nolinkurl{10.1145/3620665.3640401}}


\bibitem[Radford et~al\mbox{.}(2019)]%
        {gpt-2-radford2019language}
\bibfield{author}{\bibinfo{person}{Alec Radford}, \bibinfo{person}{Jeff Wu},
  \bibinfo{person}{Rewon Child}, \bibinfo{person}{David Luan},
  \bibinfo{person}{Dario Amodei}, {and} \bibinfo{person}{Ilya Sutskever}.}
  \bibinfo{year}{2019}\natexlab{}.
\newblock \showarticletitle{Language Models are Unsupervised Multitask
  Learners}.
\newblock  (\bibinfo{year}{2019}).
\newblock


\bibitem[Team(2026)]%
        {qwen35blog}
\bibfield{author}{\bibinfo{person}{Qwen Team}.}
  \bibinfo{year}{2026}\natexlab{}.
\newblock \bibinfo{title}{Qwen3.5: Accelerating Productivity with Native
  Multimodal Agents}.
\newblock
\urldef\tempurl%
\url{https://qwen.ai/blog?id=qwen3.5}
\showURL{%
\tempurl}


\bibitem[Wang and Oswald(2026)]%
        {wangConfidentialComputingHeterogeneous2026}
\bibfield{author}{\bibinfo{person}{Qifan Wang} {and} \bibinfo{person}{David
  Oswald}.} \bibinfo{year}{2026}\natexlab{}.
\newblock \showarticletitle{Confidential {{Computing}} on {{Heterogeneous
  CPU-GPU Systems}}: {{Survey}} and {{Future Directions}}}.
\newblock \bibinfo{journal}{\emph{Comput. Surveys}} \bibinfo{volume}{58},
  \bibinfo{number}{9} (\bibinfo{date}{Feb.} \bibinfo{year}{2026}),
  \bibinfo{pages}{230:1--230:35}.
\newblock
\showISSN{0360-0300}
\href{https://doi.org/10.1145/3793532}{doi:\nolinkurl{10.1145/3793532}}


\bibitem[Wang et~al\mbox{.}(2023)]%
        {wangHT2MLEfficientHybrid2023a}
\bibfield{author}{\bibinfo{person}{Qifan Wang}, \bibinfo{person}{Lei Zhou},
  \bibinfo{person}{Jianli Bai}, \bibinfo{person}{Yun~Sing Koh},
  \bibinfo{person}{Shujie Cui}, {and} \bibinfo{person}{Giovanni Russello}.}
  \bibinfo{year}{2023}\natexlab{}.
\newblock \showarticletitle{{{HT2ML}}: {{An}} Efficient Hybrid Framework for
  Privacy-Preserving {{Machine Learning}} Using {{HE}} and {{TEE}}}.
\newblock \bibinfo{journal}{\emph{Computers \& Security}}
  \bibinfo{volume}{135} (\bibinfo{date}{Dec.} \bibinfo{year}{2023}),
  \bibinfo{pages}{103509}.
\newblock
\showISSN{01674048}
\href{https://doi.org/10.1016/j.cose.2023.103509}{doi:\nolinkurl{10.1016/j.cose.2023.103509}}


\bibitem[Wang et~al\mbox{.}(2019)]%
        {wangScalableFullyHomomorphic2019}
\bibfield{author}{\bibinfo{person}{Wenhao Wang}, \bibinfo{person}{Yichen
  Jiang}, \bibinfo{person}{Qintao Shen}, \bibinfo{person}{Weihao Huang},
  \bibinfo{person}{Hao Chen}, \bibinfo{person}{Shuang Wang},
  \bibinfo{person}{XiaoFeng Wang}, \bibinfo{person}{Haixu Tang},
  \bibinfo{person}{Kai Chen}, \bibinfo{person}{Kristin Lauter}, {and}
  \bibinfo{person}{Dongdai Lin}.} \bibinfo{year}{2019}\natexlab{}.
\newblock \bibinfo{title}{Toward {{Scalable Fully Homomorphic Encryption
  Through Light Trusted Computing Assistance}}}.
\newblock
\showeprint[arxiv]{1905.07766}~[cs]
\href{https://doi.org/10.48550/arXiv.1905.07766}{doi:\nolinkurl{10.48550/arXiv.1905.07766}}


\bibitem[Xu et~al\mbox{.}(2025)]%
        {blb25usenixsecurity}
\bibfield{author}{\bibinfo{person}{Tianshi Xu}, \bibinfo{person}{Wen-jie Lu},
  \bibinfo{person}{Jiangrui Yu}, \bibinfo{person}{Yi Chen},
  \bibinfo{person}{Chenqi Lin}, \bibinfo{person}{Runsheng Wang}, {and}
  \bibinfo{person}{Meng Li}.} \bibinfo{year}{2025}\natexlab{}.
\newblock \showarticletitle{Breaking the layer barrier: remodeling private
  transformer inference with hybrid CKKS and MPC}. In
  \bibinfo{booktitle}{\emph{Proceedings of the 34th USENIX Conference on
  Security Symposium}} (Seattle, WA, USA) \emph{(\bibinfo{series}{SEC '25})}.
  \bibinfo{publisher}{USENIX Association}, \bibinfo{address}{USA}, Article
  \bibinfo{articleno}{137}, \bibinfo{numpages}{20}~pages.
\newblock
\showISBNx{978-1-939133-52-6}


\bibitem[Yang et~al\mbox{.}(2025)]%
        {yangQwen3TechnicalReport2025}
\bibfield{author}{\bibinfo{person}{An Yang}, \bibinfo{person}{Anfeng Li},
  \bibinfo{person}{Baosong Yang}, \bibinfo{person}{Beichen Zhang},
  \bibinfo{person}{Binyuan Hui}, \bibinfo{person}{Bo Zheng},
  \bibinfo{person}{Bowen Yu}, \bibinfo{person}{Chang Gao},
  \bibinfo{person}{Chengen Huang}, \bibinfo{person}{Chenxu Lv},
  \bibinfo{person}{Chujie Zheng}, \bibinfo{person}{Dayiheng Liu},
  \bibinfo{person}{Fan Zhou}, \bibinfo{person}{Fei Huang},
  \bibinfo{person}{Feng Hu}, \bibinfo{person}{Hao Ge}, \bibinfo{person}{Haoran
  Wei}, \bibinfo{person}{Huan Lin}, \bibinfo{person}{Jialong Tang},
  \bibinfo{person}{Jian Yang}, \bibinfo{person}{Jianhong Tu},
  \bibinfo{person}{Jianwei Zhang}, \bibinfo{person}{Jianxin Yang},
  \bibinfo{person}{Jiaxi Yang}, \bibinfo{person}{Jing Zhou},
  \bibinfo{person}{Jingren Zhou}, \bibinfo{person}{Junyang Lin},
  \bibinfo{person}{Kai Dang}, \bibinfo{person}{Keqin Bao},
  \bibinfo{person}{Kexin Yang}, \bibinfo{person}{Le Yu},
  \bibinfo{person}{Lianghao Deng}, \bibinfo{person}{Mei Li},
  \bibinfo{person}{Mingfeng Xue}, \bibinfo{person}{Mingze Li},
  \bibinfo{person}{Pei Zhang}, \bibinfo{person}{Peng Wang},
  \bibinfo{person}{Qin Zhu}, \bibinfo{person}{Rui Men}, \bibinfo{person}{Ruize
  Gao}, \bibinfo{person}{Shixuan Liu}, \bibinfo{person}{Shuang Luo},
  \bibinfo{person}{Tianhao Li}, \bibinfo{person}{Tianyi Tang},
  \bibinfo{person}{Wenbiao Yin}, \bibinfo{person}{Xingzhang Ren},
  \bibinfo{person}{Xinyu Wang}, \bibinfo{person}{Xinyu Zhang},
  \bibinfo{person}{Xuancheng Ren}, \bibinfo{person}{Yang Fan},
  \bibinfo{person}{Yang Su}, \bibinfo{person}{Yichang Zhang},
  \bibinfo{person}{Yinger Zhang}, \bibinfo{person}{Yu Wan},
  \bibinfo{person}{Yuqiong Liu}, \bibinfo{person}{Zekun Wang},
  \bibinfo{person}{Zeyu Cui}, \bibinfo{person}{Zhenru Zhang},
  \bibinfo{person}{Zhipeng Zhou}, {and} \bibinfo{person}{Zihan Qiu}.}
  \bibinfo{year}{2025}\natexlab{}.
\newblock \bibinfo{title}{Qwen3 {{Technical Report}}}.
\newblock
\showeprint[arxiv]{2505.09388}~[cs.CL]
\href{https://doi.org/10.48550/arXiv.2505.09388}{doi:\nolinkurl{10.48550/arXiv.2505.09388}}


\bibitem[Yu(2026)]%
        {yuGeeeekExplorerNanovllm2026}
\bibfield{author}{\bibinfo{person}{Xingkai Yu}.}
  \bibinfo{year}{2026}\natexlab{}.
\newblock \bibinfo{title}{{{GeeeekExplorer}}/Nano-Vllm}.
\newblock


\bibitem[Zhang et~al\mbox{.}(2025)]%
        {zhangSecureTransformerInference2025a}
\bibfield{author}{\bibinfo{person}{Jiawen Zhang}, \bibinfo{person}{Xinpeng
  Yang}, \bibinfo{person}{Lipeng He}, \bibinfo{person}{Kejia Chen},
  \bibinfo{person}{Wen-jie Lu}, \bibinfo{person}{Yinghao Wang},
  \bibinfo{person}{Xiaoyang Hou}, \bibinfo{person}{Jian Liu},
  \bibinfo{person}{Kui Ren}, {and} \bibinfo{person}{Xiaohu Yang}.}
  \bibinfo{year}{2025}\natexlab{}.
\newblock \showarticletitle{Secure {{Transformer Inference Made
  Non-interactive}}}. In \bibinfo{booktitle}{\emph{Proceedings 2025 {{Network}}
  and {{Distributed System Security Symposium}}}}. \bibinfo{publisher}{Internet
  Society}, \bibinfo{address}{San Diego, CA, USA}.
\newblock
\showISBNx{979-8-9894372-8-3}
\href{https://doi.org/10.14722/ndss.2025.230868}{doi:\nolinkurl{10.14722/ndss.2025.230868}}


\bibitem[Zhong et~al\mbox{.}(2024)]%
        {zhongDistServeDisaggregatingPrefill2024}
\bibfield{author}{\bibinfo{person}{Yinmin Zhong}, \bibinfo{person}{Shengyu
  Liu}, \bibinfo{person}{Junda Chen}, \bibinfo{person}{Jianbo Hu},
  \bibinfo{person}{Yibo Zhu}, \bibinfo{person}{Xuanzhe Liu},
  \bibinfo{person}{Xin Jin}, {and} \bibinfo{person}{Hao Zhang}.}
  \bibinfo{year}{2024}\natexlab{}.
\newblock \showarticletitle{{DistServe}: Disaggregating Prefill and Decoding
  for Goodput-optimized Large Language Model Serving}. In
  \bibinfo{booktitle}{\emph{18th USENIX Symposium on Operating Systems Design
  and Implementation}} \emph{(\bibinfo{series}{OSDI '24})}.
  \bibinfo{publisher}{USENIX Association}, \bibinfo{pages}{193--210}.
\newblock


\end{thebibliography}
\end{document}